\documentclass[journal=apcach,manuscript=article,chaptertitle=true]{achemso}

\usepackage{chemformula,bm,caption,subcaption,empheq} % Formula subscripts using \ch{}
\usepackage[T1]{fontenc} % Use modern font encodings

\usepackage{braket}
\usepackage{tikz}
\usetikzlibrary{quantikz}
\usepackage[compat=0.6]{yquant}
\usepackage{ulem}

\author{Vincent Graves}
\affiliation{National Quantum Computing Centre, Rutherford Appleton Laboratory, Harwell Campus, Didcot, Oxfordshire, OX11 0QX, UK}
\alsoaffiliation{Riverlane, St Andrews House, 59 St Andrews Street, Cambridge, CB2 3BZ,  UK}
\author{Christoph S\"underhauf}
\author{Nick S. Blunt}
\author{R\'obert Izs\'ak}
\affiliation{Riverlane, St Andrews House, 59 St Andrews Street, Cambridge, CB2 3BZ,  UK}
\email{robert.izsak@riverlane.com}
\author{Mil\'an Sz\H{o}ri}
\affiliation{Institute of Chemistry, University of Miskolc, Egyetemv\'aros A/2, H-3515 Miskolc, Hungary}

\title[An \textsf{achemso} demo]
  {The Electronic Structure of the Hydrogen Molecule: A Tutorial Exercise in Classical and Quantum Computation}

\abbreviations{RHF, UHF, FCI, strong correlation}

\begin{document}

\tableofcontents


\begin{abstract}
In this educational paper, we will discuss calculations on the hydrogen molecule both on classical and quantum computers. In the former case, we will discuss the calculation of molecular integrals that can then be used to calculate potential energy curves at the Hartree--Fock level and to correct them by obtaining the exact results for all states in the minimal basis. Some aspects of spin-symmetry will also be discussed. In the case of quantum computing, we will start out from the second-quantized Hamiltonian and qubit mappings. We then provide circuits for quantum phase estimation using two different algorithms: Trotterization and qubitization. Finally, the significance of quantum error correction will be briefly discussed. 
\end{abstract}

\section{Introduction}

It has been almost 20 years since Prof.~Csizmadia, known simply to his students as IGC, gave a series of lectures at the University of Szeged about theoretical calculations and their relevance to organic chemistry. As students (R. I., M. Sz.) attending these lectures, we knew that he had studied with Slater and was a professor of international standing who had been associated with the University of Toronto for a long time. While we had already heard of the basics of quantum mechanics and quantum chemistry, we were all eager to know more about their applications to chemical problems that we had also encountered by then in the organic chemistry lab. Who better to tell us about that than IGC, who was among the pioneers of applying Gaussian orbitals to organic molecules and was among the authors of POLYATOM\cite{barnett1963mechanized,POLYATOM,csizmadia1966non}, the first program package that could carry out such calculations? Despite his many scientific achievements, IGC never talked much about the past except to explain something to his students. He had a unique style that can be discerned from some of his writings\cite{csizmadia1991some} but that worked best in the classroom. We all remember his simple explanations of complicated mathematical subjects, usually accompanied by student-friendly illustrations that he simply called Mickey Mouse Figures. Apart from his knowledge on chemical calculation and his accessible lecturing style, his dedication set him apart from most teachers we had known: few people would have given a two-hour long lecture when struggling with a whooping cough that threatened to strangle him. In this tutorial article, we would like to pay tribute to him as an educator by providing an educational introduction to quantum chemistry methods, using the hydrogen molecule as an important first example. While IGC would have probably preferred an organic molecule and might have given less detail about the calculation than we intend to, he would have certainly approved of our using the simplest Gaussian orbital basis possible and we hope that such a simple model calculation will help the determined student to understand the machinery underlying modern quantum chemistry calculations. It is in this spirit that we offer this contribution to his memory.

\section{Theoretical Background}

\subsection{The Hartree--Fock Method}

Chemistry investigates the myriad of ways molecules may interact. With the advent of quantum mechanics, it became possible to explain these interactions in terms of those between the electrons and nuclei that make up molecules. Unfortunately, this leaves us with a large number of variables to consider if we want to describe everything that takes place in a chemist's flask. To make the problem easier to solve, several further assumptions are made beyond the axioms of quantum mechanics and special relativity needed to describe chemical systems. To start with, akin to usual practice in thermodynamics,  we may divide the universe into a system and its environment. The system is simply the part of the world we are interested in, and, for chemical purposes, this might be a number of atoms and molecules. As a first approximation, we will consider only particles in the system and neglect interactions with the environment. We will further neglect relativistic effects and the time dependence of the states of the system. Within the Born-Oppenheimer approximation, the nuclear and electronic variables are separated and the electronic problem is solved for fixed nuclear coordinates. The electronic Hamiltonian then takes the form
\begin{equation}
\hat{H}=
-\sum_{i}^{N}{\frac{1}{2}\nabla_i^2}
+\sum_{A<B}^{M}\frac{Z_A Z_B}{|\mathbf{R}_{A}-\mathbf{R}_{B}|}
-\sum_{i,A}^{N,M}\frac{Z_A}{|\mathbf{r}_{i}-\mathbf{R}_{A}|}
+\sum_{i<j}^{N}\frac{1}{|\mathbf{r}_{i}-\mathbf{r}_{j}|},
\label{Ham}
\end{equation}
where the indices $i,j$ denote electrons and $A,B$ nuclei, $\mathbf{r}_i$ is the position of an electron and $\mathbf{R}_A$ is that of a nucleus, $Z_A$ is its charge number. $N$ and $M$ are the number of electrons and nuclei, respectively. The terms in the order they appear are the kinetic energy of the electrons, the potential energy of nuclear-nuclear, nuclear-electron and electron-electron interactions.

Although the approximations so far simplify the problem considerably, the resulting quantum mechanical problem remains intractable. In the next step, the variables describing individual electrons are also separated. To fulfill the condition of antisymmetry required by the exclusion principle, an approximate many-electron wavefunction is constructed as the antisymmetrized product of functions describing a single electron
\begin{equation}
\Phi = 
\frac{1}{\sqrt{N!}}
 \begin{vmatrix}
   \phi_{1}(\mathbf{x}_1) & \phi_{2}(\mathbf{x}_1) & \cdots & \phi_{N}(\mathbf{x}_1) \\
   \phi_{1}(\mathbf{x}_2) & \phi_{2}(\mathbf{x}_2) & \cdots & \phi_{N}(\mathbf{x}_2) \\
   \vdots  & \vdots  & \ddots & \vdots  \\
   \phi_{1}(\mathbf{x}_N) & \phi_{2}(\mathbf{x}_N) & \cdots & \phi_{N}(\mathbf{x}_N) 
 \end{vmatrix}
 \label{SlaterD}
\end{equation}
The function $\Phi$ is called the Slater determinant and the one-electron functions $\phi_p$ are the spin orbitals.\cite{csizmadia1991some,mayer2003simple,szabo2012modern} Since the latter are orthonormal, the norm of $\Phi$ is also $1$. The energy of the system can then be obtained as the expectation value of of the Hamiltonian with respect to the Slater determinant,
\begin{equation}
E = \langle\Phi|\hat{H}|\Phi\rangle,
\end{equation}
implying an integration over all electronic coordinates $\mathbf{x}_p$. At this point, the spin variable of an electron ($s_p$) can also be separated from the spatial coordinates ($\mathbf{r}_p$), to yield spatial orbitals $\varphi_p$,
\begin{equation}
\phi_p(\mathbf{x}_p) = \varphi_p(\mathbf{r}_p)\sigma_p(s_p).
\label{spin-orb}
\end{equation}
For labeling the orbitals, we will use the convention that $i,j,\ldots$ refer to orbitals occupied in the Hartree--Fock ground state, $a,b,\ldots$ are unoccupied and $p,q,\ldots$ could refer to any molecular orbital. The spin-function $\sigma$ can denote a spin-up ($\alpha$) or a spin-down ($\beta$) state of a single electron identified by $s_p$. Often, the above product is denoted simply as $p\sigma$. If necessary, spin-orbital and spatial orbital labels can be distinguished by capitalizing one of them. Here, we opt for capitalizing the spin-orbital labels, which leads to the compact notation $P=p\sigma$ or $P=p\sigma_p$. Sometimes it is convenient to refer to spin orbitals in terms of their spatial component. This purpose is served by the `relative' spin notation in which the product $p\alpha$ is simply referred to as $p$ and $p\beta$ as $\bar{p}$, often combined with a simplified notation for the particle label, as in $p(1)$ or $\varphi_p(1)$ denoting $\varphi_p(\mathbf{r}_1)\alpha(s_1)$. For closed shell systems, with an even number of paired electrons, it is convenient to assume that spin-orbitals of a spin-up/spin-down pair have identical spatial parts. This leads to the restricted Hartree-Fock (RHF) theory we will be mostly concerned with in the remainder of this section. For open shell systems, the choice arises whether spin symmetry is enforced as in restricted open-shell Hartree-Fock (ROHF), or whether the spin-up and spin-down electrons are allowed to have different spatial parts thereby breaking spin symmetry but providing more flexibility in the optimization process as in unrestricted Hartree-Fock (UHF) theory. We will also come across examples of such states in later parts of this tutorial.

Using the fact that the spin functions $\alpha$ and $\beta$ are orthonormal, the following expression can be obtained for the RHF energy using the Slater-Condon rules\cite{mayer2003simple}
\begin{equation}
E_{\text{RHF}} = \langle \Phi |\hat{H}|\Phi \rangle = E_n + 2\sum_i^{N_o} (i|\hat{h}|i) + 2\sum_{ij}^{N_o}(ii|jj)-\sum_{ij}^{N_o}(ij|ij),
\label{HFE}
\end{equation}
where $E_n$ is simply the nuclear-nuclear interaction term in Eq.~\eqref{Ham}. $N_o$ is the number of occupied spatial orbitals which for closed shell systems is half the number of electrons, $N_o=N/2$, simply because each spatial orbital can be used to construct two spin-orbitals via Eq.~\eqref{spin-orb}. The one electron integral reads
\begin{equation}
(p|\hat{h}|q) = \int \varphi_p^*(\mathbf{r}) \hat{h}(\mathbf{r}) \varphi_q(\mathbf{r})\,\text{d}\mathbf{r},
\end{equation}
with $\hat{h}$ containing the kinetic energy term of the electrons and the nuclear-electron interaction energy in Eq.~\eqref{Ham}. Finally, the two-electron term is
\begin{equation}
(pq|rs) = \iint \frac{\varphi^*_p(\mathbf{r}_1)\varphi_q(\mathbf{r}_1)\varphi^*_r(\mathbf{r}_2)\varphi_s(\mathbf{r}_2)}{|\mathbf{r}_1-\mathbf{r}_2|}\,\text{d}\mathbf{r}_1\text{d}\mathbf{r}_2,
\end{equation}
and represents the remaining electron-electron interaction term in Eq.~\eqref{Ham}. Note that these integrals are defined in terms of spatial orbitals but the spin-orbital equivalents are easily defined as $(P|\hat{h}|Q)=(p|\hat{h}|q)\delta_{\sigma_p\sigma_q}$ and $(PQ|RS)=(pq|rs)\delta_{\sigma_p\sigma_q}\delta_{\sigma_r\sigma_s}$. Here, and in the rest of the paper, integration over electronic coordinates implies an integration over the entire physical space. See the Supporting Information for further details on integrals.

To find the lowest-energy determinant $\Phi$, the energy needs to be minimized under the constraint that $\Phi$ is normalized. This is a quite involved task in general, and in a final approximation,
the molecular orbitals (MO) $\varphi_p$ are expanded in terms of known atomic orbitals (AO), $\chi_\mu(\mathbf{r})$, serving as basis functions,
\begin{equation}
\varphi_p = \sum_{\mu}^{N_b} C_{\mu p} \chi_{\mu}.
\end{equation}
Here $C_{\mu p}$ is an element of the MO coefficient matrix $\mathbf{C}$ and $N_b$ is the number of basis functions such that $N_o\leq N_b$. This yields the algebraic form of the Hartree--Fock equations, sometimes called the Hartree--Fock-Roothaan-Hall equations,\cite{csizmadia1991some,szabo2012modern,helgaker2014molecular}
\begin{equation}
\mathbf{FC} = \mathbf{SCE},
\end{equation}
with the elements of the Fock matrix $\mathbf{F}$ and the overlap matrix $\mathbf{S}$ defined as
\begin{equation}
F_{\mu\nu} = \int\chi_{\mu}(\mathbf{r})\hat{f}(\mathbf{r})\chi_{\nu}(\mathbf{r})\,\text{d}\mathbf{r},\quad
S_{\mu\nu} = \int\chi_{\mu}(\mathbf{r})\chi_{\nu}(\mathbf{r})\,\text{d}\mathbf{r},
\label{SandF}
\end{equation}
and $\mathbf{E}$ being the diagonal matrix containing the molecular orbital energies. Note that from this point on we will assume that quantities are real and will not denote complex conjugation any more. As the Fock operator
\begin{equation}
\hat{f}[\{\varphi_i\}](\mathbf{r}_1) = \hat{h}(\mathbf{r}_1) + \sum_j^{N_o}\int\varphi_j(\mathbf{r}_2;\mathbf{R})\frac{2-\hat{P}_{12}}{|\mathbf{r}_1-\mathbf{r}_2|}\varphi_j(\mathbf{r}_2;\mathbf{R})\;\text{d}\mathbf{r}_2,
\end{equation}
itself depends on the orbitals that we seek to optimize, this is a \emph{self-consistent} eigenvalue problem, the solutions of which must be found in an iterative manner. The operator $\hat{P}_{12}$ swaps the coordinate labels to account for antisymmetry. The resulting Fock matrix has the general form
\begin{equation}
F_{\mu\nu} = h_{\mu\nu} + G_{\mu\nu},
\label{Fock}
\end{equation}
where the core term $h_{\mu\nu}$ itself consists of two contributions, 
\begin{equation}
h_{\mu\nu} = T_{\mu\nu} + V_{\mu\nu},
\end{equation}
with
\begin{equation}
T_{\mu\nu} = -\frac{1}{2}\int\chi_{\mu}(\mathbf{r})\nabla^2 \chi_{\nu}(\mathbf{r})\,\text{d}\mathbf{r},\quad
\end{equation}
and
\begin{equation}
V_{\mu\nu} = \sum_{A}^M V_{\mu\nu}(A),\quad
V_{\mu\nu}(A) = -Z_A\int\frac{\chi_{\mu}(\mathbf{r})\chi_{\nu}(\mathbf{r})}{|\mathbf{r}-\mathbf{R}_{A}|}\,\text{d}\mathbf{r},
\end{equation}
while the electronic interaction term consists of a direct Coulomb term and an exchange term,
\begin{equation}
G_{\mu\nu} = \sum_{\kappa\lambda}^{N_b}P_{\kappa\lambda} (\mu\nu|\kappa\lambda)
-\frac{1}{2} \sum_{\kappa\lambda}^{N_b}P_{\kappa\lambda} (\mu\kappa|\nu\lambda)
\end{equation}
with
\begin{equation}
(\mu\nu|\kappa\lambda) = \iint \frac{\chi_{\mu}(\mathbf{r}_1)\chi_{\nu}(\mathbf{r}_1)\chi_{\kappa}(\mathbf{r}_2)\chi_{\lambda}(\mathbf{r}_2)}{|\mathbf{r}_1-\mathbf{r}_2|}\,\text{d}\mathbf{r}_1\text{d}\mathbf{r}_2,
\end{equation}
where the charge-density matrix is defined as
\begin{equation}
P_{\kappa\lambda} = 2\sum_i^{N_o} C_{\kappa i} C_{\lambda i}.
\end{equation}
Finally, the spin-restricted Hartree--Fock energy can be written in terms of the AO quantities as
\begin{equation}
E_{\text{RHF}} = E_{n} + \frac{1}{2}\sum_{\mu\nu}^{N_b}P_{\mu\nu}(h_{\mu\nu}+F_{\mu\nu}).
\label{TrE}
\end{equation}

\subsection{Electron Correlation}

The Hartree--Fock solution has several deficiencies that originate in the approximations made. Over the decades several methods have been devised that improve one or more of these approximations, starting from the Hartree--Fock solution.\cite{szabo2012modern,helgaker2014molecular,bartlett1994applications,knowles2000ab} Other than the choice of the AO basis, improving on the treatment of interelectronic interactions in the Hartree--Fock method is the most important issue in practical calculations. In particular, in the Hartree--Fock model, the electrons with parallel and anti-parallel spins are treated differently in that the probability of finding two electrons at the same place is zero in the former case (presence of a Fermi hole) and non-zero in the latter (lack of a Coulomb hole). This is a consequence of representing the many-body wavefunction using a single Slater determinant. However, the manifold of Slater determinants, $\{\Phi_I\}$, obtained by replacing the occupied orbitals in Eq.~\eqref{SlaterD} with virtual ones in all possible ways, can be used to build an improved wavefunction $\Psi$ as a linear combination
\begin{equation}
\Psi = \sum_I^{N_\Phi} \mathcal{C}_I\Phi_I.
\label{FCI}
\end{equation}
If the expansion contains all $N_\Phi$ possible Slater determinants in a given basis, then this full configuration interaction (FCI) expansion will yield the exact solution in that basis if the normalized coefficients $C_I$ are optimised. To find these, the following eigenproblem must be solved
\begin{equation}
\mathbf{H}\bm{\mathcal{C}}=\mathcal{E}\bm{\mathcal{C}},
\end{equation}
where $\mathbf{H}$ is the matrix representation of the Hamiltonian with elements
\begin{equation}
H_{IJ} = \langle\Phi_I | \hat{H} | \Phi_J\rangle.
\label{HMat}
\end{equation}
The correlation energy $E_c$ is then the difference between the exact energy $\mathcal{E}$ and the Hartree--Fock energy $E$,
\begin{equation}
E_c = \mathcal{E} - E.
\end{equation}
However, determinants are not the only basis in which $\Psi$ can be expanded. Unlike determinants, configuration state functions (CSF)\cite{helgaker2014molecular} are eigenfunctions of the total spin squared operator and can be represented as a sum of $N_C$ determinants,
\begin{equation}
\Theta = \sum_I^{N_C} D_I\Phi_I,
\label{DefCSF}
\end{equation}
where $D_I$ are fixed coefficients and $\Theta$ is a CSF. For a given spin sublevel, there are in fact usually fewer CSFs than there are determinants. Using CSFs instead of determinants may change how many basis states have large coefficients in the FCI expansion, as does rotating the orbitals between occupied and virtual spaces.\cite{izsak2023measuring} If $N_\Theta$ is the number of CSFs in an FCI expansion, then in practice this means that $N_\Theta\leq N_\Phi$. In the closed shell case, the issue does not arise because the closed shell RHF determinant is already a CSF and that covers many important chemical cases, including most organic compounds. The use of CSFs is especially advantageous for open-shell systems where the difference between $N_\Theta$ and $N_\Phi$ can be quite substantial apart from high spin cases where the CSF is again a single determinant. The ROHF approach starts, when possible,\cite{izsak2023measuring} from a single CSF rather than a single determinant as it is done in the UHF approach. It should also be noted that the more wide-spread use of CSFs is hindered by the fact that the ROHF equations can be difficult to converge in some cases and in correlation methods they are quite difficult to implement for popular methods like coupled cluster theory.

\section{The Hydrogen Molecule in a Minimal Basis}

\subsection{Possible States and Their Long Range Behaviour} \label{Sec:states}

Consider a single H$_2$ molecule and let $\chi_\mu$ be an atomic orbital\cite{csizmadia1991some} (AO) on one of the H atoms, and $\chi_\nu$ another AO on the other H atom. Because this simple problem has only two basis functions and is highly symmetric, it is possible to describe some of its properties, especially at large internuclear separations, even without solving the HF and FCI equations. In this section, we will discuss such general considerations, then move on to the actual calculations.

Due to the symmetries of H$_2$, we know that there are only two possibilities of combining $\chi_\mu$ and $\chi_\nu$ into molecular orbitals. The MO coefficients must have the same magnitude with the same or opposite signs. After normalization, this yields the bonding orbital $\varphi_i$
\begin{equation}
\varphi_i = \frac{1}{\sqrt{2(1+S_{\mu\nu})}}(\chi_\mu + \chi_\nu),
\label{phii}
\end{equation}
while the anti-bonding orbital $\varphi_a$ has the form
\begin{equation}
\varphi_a = \frac{1}{\sqrt{2(1-S_{\mu\nu})}}(\chi_\mu - \chi_\nu).
\label{phia}
\end{equation}
In terms of the MO coefficient matrix $\mathbf{C}$, this means that
\begin{equation}
\mathbf{C} = 
\begin{pmatrix}
\frac{1}{\sqrt{2(1+S_{\mu\nu})}} & \frac{1}{\sqrt{2(1-S_{\mu\nu})}} \\
\frac{1}{\sqrt{2(1+S_{\mu\nu})}} & -\frac{1}{\sqrt{2(1-S_{\mu\nu})}} \\
\end{pmatrix}.
\label{CMat}
\end{equation}
For the purposes of analysing long range behaviour, the AOs can be assumed to be orthonormal, since the two AOs barely overlap at large bond lengths.  Thus, for the remainder of this section, we will assume that  $S_{\mu\nu}\approx 0$. The lowest-energy RHF determinant is then 
\begin{equation}
\Phi_0 = |i\bar{i}|,
\label{det0}
\end{equation}
where
\begin{equation}
|i\bar{i}|\equiv |\varphi_i\bar{\varphi}_i| = 
\frac{1}{\sqrt{2}}(\varphi_i(1)\bar{\varphi}_i(2) - \varphi_i(2)\bar{\varphi}_i(1)).
\end{equation}
Remember that $\varphi_i$ and $\bar{\varphi}_i$ now denote spin-orbitals in the relative spin notation. It should also be remembered that $\Phi_0=|i\bar{i}|$ is one of six possible determinants that can be constructed from the orbitals $i$ and $a$ spanning the Hilbert space, the other five being $|a\bar{a}|$, $|ia|$, $|i\bar{a}|$, $|\bar{i}a|$ and $|\bar{i}\bar{a}|$. Substituting Eq.~\eqref{phii}, one gets
\begin{equation}
\Phi_0 = \Phi_{C0} + \Phi_{I0},
\end{equation}
where $\Phi_{0}$ consists of a covalent part
\begin{equation}
\Phi_{C0} = \frac{1}{2}(|\mu\bar{\nu}| + |\nu\bar{\mu}|),
\end{equation}
and an ionic part
\begin{equation}
\Phi_{I0} = \frac{1}{2}(|\mu\bar{\mu}| + |\nu\bar{\nu}|).
\end{equation}
The covalent contribution consists of AO basis determinants in which one electron is assigned to one H atom via $\mu$ and the other electron to the other H atom via $\nu$, which is what is expected in a homolytic dissociation process. The ionic contribution on the other hand consists of AO determinants in which both electrons are assigned to one atom only. The fact that in $\Phi_0$ the two contributions come with an equal weight leads to what is known as the dissociation catastrophe of Hartree--Fock theory.\cite{bartlett1994applications,knowles2000ab} Since $\Phi_{C0}$ describes a homolytic and $\Phi_{I0}$ a heterolytic process and since the latter requires a much higher energy, the total dissociation curve produces an artificially large dissociation energy for the homolytic process. The customary solution is to construct another one of the six Hilbert space basis states, the doubly excited determinant
\begin{equation}
\Phi_1 = |a\bar{a}|,
\end{equation}
which, after following a similar procedure as above, is found to be
\begin{equation}
\Phi_1 = -\Phi_{C0} + \Phi_{I0}.
\end{equation}
We may now define a two-determinant trial wavefunction
\begin{equation}
\Psi_0 = \mathcal{C}_0\Phi_0 + \mathcal{C}_1\Phi_1,
\end{equation}
which can be written as
\begin{equation}
\Psi_0 = (\mathcal{C}_0-\mathcal{C}_1)\Phi_{C0} + (\mathcal{C}_0+\mathcal{C}_1)\Phi_{I0}.
\label{VB}
\end{equation}
It is clear that if $\mathcal{C}_0=-\mathcal{C}_1$, the ionic contribution vanishes and the covalent contribution survives. This trial function has the necessary flexibility to describe the entire curve in a qualitatively correct way: close to the equilibrium $\mathcal{C}_0\approx 1$, which agrees well with the fact that HF is a good description of the H$_2$ molecule at equilibrium distance. This analysis is also identical with the result obtained from a valance bond (VB) construction of the wavefunction. Similar results can be obtained for the correct heterolytic curve starting from the doubly excited determinant $\Phi_1$.

So far we have only considered the ground state and the doubly excited state within the minimal basis. When it comes to singly excited states, it is useful to represent them using CSFs, i.e., linear combinations of determinants that are spin-eigenstates, as mentioned above. For a singlet state, this has the form
\begin{equation}
\Theta_S = \frac{1}{\sqrt{2}}(|i\bar{a}| - |\bar{i}a|),
\end{equation}
Thus, this state is described by a single CSF ($N_\Theta = 1$) consisting of two determinants ($N_\Phi = 2$), providing an example of our discussion after Eq.~\eqref{DefCSF}. The two determinants used here can also be analyzed in terms of AO basis determinants,
\begin{equation}
|i\bar{a}| = -\Phi_{C1} + \Phi_{I1},
\end{equation}
\begin{equation}
|a\bar{i}| = \Phi_{C1} + \Phi_{I1},
\end{equation}
where the ionic and covalent contributions are
\begin{equation}
\Phi_{C1} = \frac{1}{2}(|\mu\bar{\nu}| - |\nu\bar{\mu}|),
\end{equation}
\begin{equation}
\Phi_{I1} = \frac{1}{2}(|\mu\bar{\mu}| - |\nu\bar{\nu}|).
\end{equation}
Therefore,
\begin{equation}
\Theta_S = \sqrt{2}\Phi_{I1},
\label{deftheta}
\end{equation}
which means that this is a fully ionic solution at long distance. Similarly, the three degenerate triplet states,
\begin{equation}
\Phi_{T}^{+} = |ia|=-|\mu\nu|,
\end{equation}
\begin{equation}
\Theta_T = \frac{1}{\sqrt{2}}(|i\bar{a}| + |\bar{i}a|)=\sqrt{2}\Phi_{C1},
\end{equation}
\begin{equation}
\Phi_T^{-} = |\bar{i}\bar{a}|=-|\bar{\mu}\bar{\nu}|,
\end{equation}
all of which have a covalent character. 

Next, we will turn to the calculation of the electronic energies for all these various states at different levels of theory. This requires the construction of the integrals in Eq.~\eqref{SandF}. The simplest model that can be evaluated without the aid of a computer assumes that the basis functions $\chi_{\mu}$ and $\chi_\nu$ are simple normalized Gaussians\cite{csizmadia1991some}
\begin{equation}
\chi_{\mu} = \left(\frac{2\alpha}{\pi}\right)^{\frac{3}{4}}e^{-\alpha(\mathbf{r}+\mathbf{R})^2},
\quad
\chi_{\nu} = \left(\frac{2\alpha}{\pi}\right)^{\frac{3}{4}}e^{-\alpha(\mathbf{r}-\mathbf{R})^2},
\end{equation}
where we have assumed that the two atoms are at an equal distance $\mathbf{R}$ from the origin along the z-axis, i.e., that $\mathbf{R}=(0,0,R)$. The orbital exponent $\alpha = \frac{8}{9\pi}$ can be found following a procedure outlined in the Supporting Information where all the necessary integrals with details of their evaluation (for both Gaussian and Slater type orbitals) are also given.

\subsection{The RHF/ROHF Potential Energy Curves}

The energy expression in Eq.~\eqref{HFE} is particularly simple for the lowest energy determinant $\Phi_0$ (sometimes referred to as the RHF ground state) of H$_2$,
\begin{equation}
E_0 = \langle\Phi_0|\hat{H}|\Phi_0\rangle = E_n + 2 (i|\hat{h}|i) + (ii|ii).
\label{E0RHF}
\end{equation}
Once the intergrals are constructed and an initial guess of $\mathbf{C}$ is found, the next step should be to build the Fock matrix and optimize $P_{\mu\nu}$ iteratively. Fortunately, the symmetry adapted orbitals in Eq.~\eqref{phii} and Eq.~\eqref{phia} turn out to be the self-consistent solutions of the Hartree--Fock equations. For these orbitals, the charge density matrix is
\begin{equation}
\mathbf{P} = \frac{1}{1+S_{\mu\nu}}
\begin{pmatrix}
\;1 & \;\;1\; \\
\;1 & \;\;1\;
\end{pmatrix}.
\end{equation}
With this, the Fock matrix can be built as in Eq.~\eqref{Fock}, while the energy can be obtained as in Eq.~\eqref{TrE}. The explicit construction of these quantites and the final analytic formulae are discussed in more detail in the Supporting Information. 

To obtain the doubly excited state $\Phi_1$, Eq.~\eqref{HMat} should be evaluated. Fortunately, for H$_2$ in the minimal basis, a simpler route is available by simply relabeling all $i$ to $a$ in the energy formula,
\begin{equation}
E_1 =  \langle\Phi_1|\hat{H}|\Phi_1\rangle = E_n + 2 (a|\hat{h}|a) + (aa|aa).
\end{equation}
This amounts to constructing a new density,
\begin{equation}
\bar{\mathbf{P}} = \frac{1}{1-S_{\mu\nu}}
\begin{pmatrix}
1 & -1 \\
-1 & 1 \\
\end{pmatrix},
\end{equation}
which then produces a modified Fock matrix. The procedure from this point on is very similar to the case of $E_0$ and is detailed in the Supporting Information.

\begin{figure}[ht]
\includegraphics[scale=1.5]{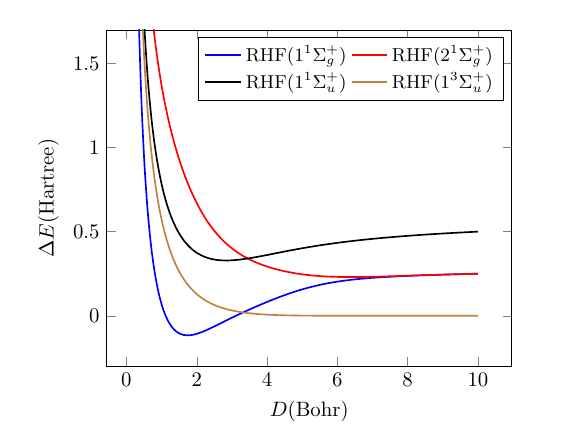}
\caption{The RHF potential energy curves for H$_2$ in a minimal basis. There are three singlet and three degenerate triplet states.}
\label{fig:RHF}
\end{figure}

Finally, the HF energies, $E_S$ and $E_T$, of the singly excited singlet and triplet states can be obtained from Eq.~\eqref{HMat}, by using the Slater-Condon rules,\cite{mayer2003simple}
\begin{equation}
E_S = \langle\Theta_S|\hat{H}|\Theta_S\rangle = E_n + (i|\hat{h}|i) + (a|\hat{h}|a) + (ii|aa) + (ia|ia),
\end{equation}
and
\begin{equation}
E_T = \langle\Theta_T|\hat{H}|\Theta_T\rangle = E_n + (i|\hat{h}|i) + (a|\hat{h}|a) + (ii|aa) - (ia|ia).
\end{equation}
The AO expressions and the final analytical formulae are again given in the Supporting Information.

Fig.~\ref{fig:RHF} displays the dissociation curves $\Delta E_{\text{X}} = E_{\text{X}} - 2E_{\text{H}}$ for all the possible Hartree--Fock states in the minimal basis. Thus, $E_{\text{X}}$ can be the RHF energy of the $1{^1\Sigma^+_g}$ singlet ground state ($E_0$), the doubly excited $2{^1\Sigma^+_g}$ singlet state ($E_1$), the singly excited $1{^1\Sigma^+_u}$ singlet state ($E_S$) and one of the degenerate $1{^3\Sigma^+_u}$ triplet states ($E_T$), while $E_{\text{H}}$ is the energy of the H atom. Around the equilibrium distance, all curves behave reasonably, the ground state and the singly excited state have a minimum indicating a stable structure for H$_2$ in these states. As the two H atoms are pulled apart, the ground state and the doubly excited states converge. From the formulae provided in the Supporting Information, it is easily seen that $\Delta E_0$ and $\Delta E_1$ both converge to the value $\sqrt{\alpha/\pi}\, E_\text{h}$ as the internuclear distance $D$ goes to infinity, while $\Delta E_S$ approaches $2\sqrt{\alpha/\pi}\, E_\text{h}$ and $\Delta E_T$ goes to 0 as $D\to\infty$. In principle, both the lowest energy singlet ($\Delta E_0$) and triplet ($\Delta E_T$) curves should converge to zero at infinite separation since they both are expected to yield two neutral H atoms with antiparallel (singlet) or parallel (triplet) spins. These should be identical in energy in the absence of an external magnetic field. The fact that the ground state RHF curve in particular does not approach zero is often referred to as the `dissociation catastrophe' of the RHF method.\cite{bartlett1994applications,knowles2000ab} It shows that RHF does not produce two H atoms at infinite distance, but due to the weight of the ionic contributions mentioned in Sec.~\ref{Sec:states}, it significantly overshoots, although it should be noted that it is still well under the purely ionic limit at $2\sqrt{\alpha/\pi}\, E_\text{h}$. One way to solve this problem is to mix various states of the same spin and spatial symmetry; we will consider this approach in the next section.

\subsection{The FCI Potential Energy Curves}

To overcome the problems of the RHF method, the wavefunction can also be expanded as in Eq.~\eqref{FCI}. This means that the matrix Hamiltonian in the basis of many-electron basis states shown in Eq.~\eqref{HMat} must be diagonalized. Notice that neither of the singlet RHF states mix with the triplet as they have different spin symmetry and $\Theta_S$ does not mix with $\Phi_0$ or $\Phi_1$, as they have they have different spatial symmetry. Thus, the only non-zero off-diagonal elements in Eq.~\eqref{HMat} are those between $\Phi_0$ and $\Phi_1$, yielding a conveniently simple two-by-two matrix
\begin{equation}
\mathbf{H}=
\begin{pmatrix}
E_0 & g \\
g & E_1 \\
\end{pmatrix},
\end{equation}
where $g = \langle\Phi_1|\hat{H}|\Phi_0\rangle = (ia|ia)$, discussed more explicitly in the Supporting Information. 

\begin{figure}[ht]
\includegraphics[scale=1.5]{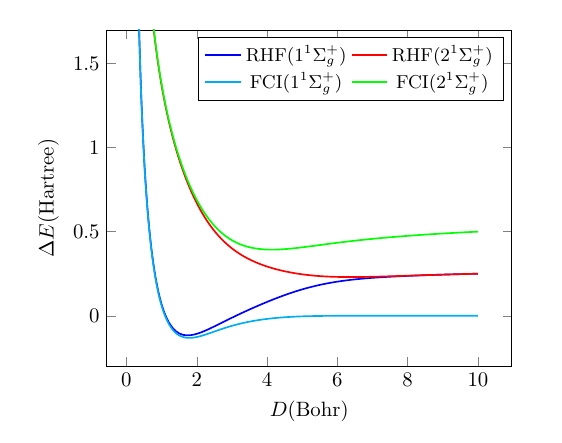}
\caption{The FCI and the corresponding RHF potential energy curves.}
\label{fig:FCI}
\end{figure}

As discussed before in Sec.~\ref{Sec:states}, the mixture of $\Phi_0$ or $\Phi_1$ is enough to produce the correct ground-state solution in the minimal basis, due to the cancellation of ionic terms. The eigenvalues of $H$, shown in Fig.~\ref{fig:FCI}, are
\begin{equation}
E_{\pm} = A\pm\frac{\sqrt{\omega^2+4g^2}}{2},\quad
A = \frac{1}{2}(E_0 + E_1),\quad
\omega = E_1-E_0,
\end{equation}
where $A$ is the average RHF energy of the two states, while $\omega$ is the excitation energy. Using the formulae of the Supporting Information, it is now easy to show that the FCI solution with the minus sign, $E_-$ approaches 0 as $D\to\infty$ corresponding to the correct covalent dissociation limit. Furthermore, the other solution, $E_+$, converges to the correct ionic limit $2\sqrt{\alpha/\pi}\, E_\text{h}$. Thus, within the minimal basis, only the triplet and the singly excited singlet states are described correctly at the RHF level, it is necessary to mix two RHF states to recover the exact solutions for the other two. As we will see in the next section, there is an alternative: breaking the spin symmetry also removes the dissociation catastrophe.

\subsection{The UHF Potential Energy Curves}

The RHF solution in Eq.~\eqref{CMat} has the property that $C_{\mu i} = C_{\nu i}$ and for the occupied MO $i$. The unrestricted Hartree--Fock model differs from RHF in that there are two different sets of spatial orbitals for electrons with alpha and beta spins, $i_\alpha$ and $i_\beta$. Since these MOs span the same space as the RHF solution, we may represent them using the RHF orbitals as a basis,\cite{szabo2012modern}
\begin{equation}
\varphi_{i_\alpha} = U^\alpha_{ii}\varphi_{i} + U^\alpha_{ia}\varphi_a,
\label{Utrafo_i_a}
\end{equation}
\begin{equation}
\varphi_{i_\beta} = U^\beta_{ii}\varphi_{i} + U^\beta_{ia}\varphi_a.
\label{Utrafo_i_b}
\end{equation}
with $U^\alpha$ and $U^\beta$ being the unitary transformations that yield $i_\alpha$ and $i_\beta$. In this case, the MO coefficients belonging to the two AOs are not fixed by symmetry and need not have the same magnitude, i.e., $C_{\mu i_\sigma} \neq C_{\nu i_\sigma}$, for $\sigma=\alpha, \beta$. Note also that the choice $U^\alpha = U^\beta$, i.e., obtaining a single set of rotated orbitals is what is done in orbital localization procedures. We will comment on this choice later on. On the other hand, while in UHF $U^\alpha \neq U^\beta$, the number of alpha and beta electrons are equal which is reflected in the solution: both alpha and beta orbitals can be obtained from the RHF ones by a rotation of the same angle but opposite direction, i.e., $U^\alpha=U^{\beta\dagger}$, $U^\alpha_{ii}=U^\beta_{ii}\equiv U_{ii}$ and $U^\alpha_{ia}=-U^\beta_{ia}\equiv U_{ia}$. On substitution into the UHF determinant
\begin{equation}
\Phi_{\text{UHF}} = |i_{\alpha}\bar{i}_{\beta}| = U_{ii}^2 \Phi_0 - U_{ia}^2\Phi_1 -
\sqrt{2}U_{ii}U_{ia}\Theta_T,
\end{equation}
which reveals the spin-symmetry-broken nature of the UHF wavefunction since it mixes singlet and triplet states. Evaluating the energy as an expectation value gives
\begin{equation}
E_{\text{UHF}} = 
U_{ii}^4 (E_0 + E_1 +2g -2E_T) +2U^2_{ii}(E_T -E_1 - g) + E_1,
\end{equation}
where the normalization condition was used to eliminate $U_{ia}$. This expression can be minimized as a function of $U_{ii}$ with the result
\begin{equation}
U_{ii}^2 = \frac{E_1 + g -E_T}{E_0 + E_1 +2g -2E_T}.
\end{equation}
Note that because of normalization, it must be true that $U_{ii}^2\leq 1$ ($U_{ii}$ is the cosine of the rotation angle). It turns out that the above function decreases monotonically as a function of $D$ and it approaches 0.5 at infinity. Thus, we need only find out where it takes the value $U_{ii}=1$, which is the case if
\begin{equation}
E_0 +g -E_T = 0.
\end{equation}
This happens at the Coulson-Fischer point at a distance of $D_{CF} = 2.4653\, \text{Bohr}$.
Thus, the optimized UHF energy for the ground state becomes
\begin{equation}
E_{\text{UHF}} = 
\begin{cases}
E_0, & D<D_{CF},  \\
E_1 - \frac{(E_1+g-E_T)^2}{E_0+E_1+2g-2E_T}, & D_{CF}\leq D.
\end{cases}
\end{equation}
It is worth mentioning at this point that choosing $U^\alpha = U^\beta$ and carrying out a similar optimization would not have improved the ground state energy but rather flipped the orbitals and converged to the doubly excited state solution.

\begin{figure}[ht]
\includegraphics[scale=1.5]{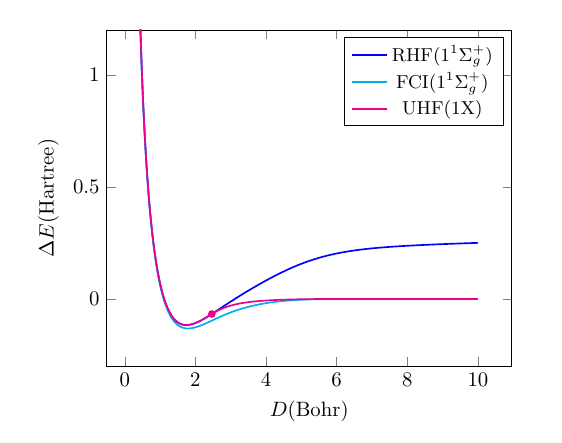}
\caption{The RHF, UHF and FCI ground-state potential energy curves. The Coulson-Fischer point is indicated by a circle on the UHF curve.}
\label{fig:UHF}
\end{figure}

Fig.~\ref{fig:UHF} shows the RHF, UHF and FCI ground-state solutions. Unlike the RHF solution, the UHF curve indeed approaches the FCI limit at infinite distance. The UHF solution is often a convenient starting point for electron correlation methods since it is a much more flexible reference point than the spin-restricted alternative, which can often only achieve a qualitatively correct starting point by mixing several determinants or CSFs. The fact that it breaks spin symmetry is undesirable because in molecular systems (with finite size and number of electrons), the exact wavefunction does not break the symmetries\cite{gross1996role} of the Hamiltonian only approximation wavefunctions do. Conversely, the amount of spin-contamination in the UHF state (in this case the amount of triplet mixing in $\Phi_{\text{UHF}}$) can be regarded as a measure of deviation from the exact state, or, sometimes strong correlation associated with symmetry breaking. Nevertheless, while the spin-symmetry-broken character of UHF can also be problematic,\cite{bartlett1994applications,knowles2000ab,szabo2012modern} the FCI solution in Eq.~\eqref{FCI} is much harder to obtain. Consequently, many approximate approaches have been developed on classical computers to tackle this problem, and more recently the potential benefit of quantum computers in solving this problem has also been investigated. In the next section, we will continue the discussion of the H$_2$ molecule from the perspective of quantum computers. In doing so, we will provide an introduction to the topic of quantum algorithms for solving quantum chemistry problems, which is currently an active and growing area of research.

\section{The Hydrogen Molecule on the Quantum Computer}

\subsection{Chemistry and Quantum Computing}

Chemistry is regarded as the application area which will be the first to benefit from quantum computing. At the core of this expectation is a potential exponential speed-up with respect to classical calculations. Quantum computers consist of a series of qubits, i.e., two level systems that can be characterised as an arbitrary superposition $|m\rangle$ between the basis states $|0\rangle$ and $|1\rangle$,
\begin{equation}
|m\rangle = \alpha_m |0\rangle + \beta_m |1\rangle,
\end{equation}
where $\alpha_m$ and $\beta_m$ are complex coefficients obeying the normalization condition $|\alpha_m|^2 + |\beta_m|^2 = 1$. The basis states $|0\rangle$ and $|1\rangle$ are known as the spin-up and spin-down states or $\alpha$ and $\beta$ spins in chemistry and they can be represented as two dimensional unit vectors. To carry out computations on qubits, appropriate quantum gates are introduced as various unitary operators. The simplest of these are representable as the two-by-two Pauli spin matrices. As a result of the action of various gates on such qubits, they end up in a state that could be characterized by the coefficients $\alpha_m$ and $\beta_m$ using the basis states $|0\rangle$ and $|1\rangle$ and which could also be measured, although at the cost of the collapse of the qubit into one of the basis states. To better understand the advantage with respect to classical computing, let us consider the two qubit state $|mn\rangle$ (understood here as the tensor product state $|m\rangle\otimes |n\rangle$) and its expansion in terms of basis states,
\begin{equation}
|mn\rangle = 
\alpha_m\alpha_n |00\rangle +
\alpha_m\beta_n |01\rangle +
\beta_m\alpha_n |10\rangle +
\beta_m\beta_n |11\rangle.
\end{equation}
To characterize this state, we need two qubits on a quantum computer whereas classically we would need to store four coefficients corresponding to the two-bit (tensor product) basis states $|00\rangle$, $|01\rangle$, $|10\rangle$ and $|11\rangle$. In general, an $N$ qubit state would correspond to a superposition of length $2^N$. To see how much of this applies to chemistry, we will need a mapping that relates the operators to quantum gates and determinants to the qubit basis states.  We will consider such a mapping in more detail in Sec~\ref{SSec:SQM} after introducing second quantization as an intermediate step to arrive at this mapping in Sec.~\ref{SSec:2QH}. For now, we will only note that the size of the FCI expansion in Eq.~\eqref{FCI} grows exponentially with the system size and the hope is to reduce this to linear scaling in the number of qubits by exploiting the quantum properties of qubits in quantum computers. This would make classically intractable problems like the FeMoco of nitrogenase more easily approachable with quantum computers. There are of course several caveats to this, for example various symmetries could reduce costs both on classical and quantum computers, or the principle of locality might do likewise in cases with no obvious symmetries. The question of exponential speed-up for chemistry in practical calculations is a complicated one that has been scrutinized in more detail elsewhere \cite{Lee_2023}.

When it comes to chemical applications involving quantum computing, it is also useful to distinguish between three categories discussed in the literature:
\begin{itemize}
\item{
Calculations on quantum hardware: 
These are at a very early stage, nothing of any far-reaching consequence for chemistry is possible at the moment.  It is nevertheless possible to carry out very simple chemical calculations, like to obtain the ground state energy of the H$_2$ molecule in the minimal basis. These are important milestones for hardware development.\cite{graham_2022, blunt_2023,yamamoto2025quantum}
}
\item{
Calculations using classical simulators:
Nothing that would outperform classical methods is possible (since the simulation is on classical hardware). These calculations are carried out for testing quantum algorithms at reasonable sizes (i.e., larger than H$_2$). We have discussed them elsewhere\cite{izsak2023quantum} and they are not pertinent to our discussion here.
}
\item{
Resource estimation:
No productive calculations are carried out either on quantum or classical hardware, but the resource requirements of calculations in terms of the number of qubits required and the number of gates contained in quantum circuits are calculated for a given Hamiltonian. This then can be used to estimate the time it would take to carry out a calculation to a given accuracy \emph{assuming} the appropriate hardware were available. Classically intractable cases are targeted in such studies, e.g., the FeMoco of nitrogenase\cite{Reiher_2017,low2025fast}. Such studies demonstrate the potential usefulness of quantum computing for chemistry and describe the hardware we would need to carry out such calculations.\cite{Reiher_2017,bluntPerspectiveCurrentStateoftheArt2022,low2025fast}
}
\end{itemize}
It will also be helpful to follow the literature in distinguishing between various ``eras'' of quantum computing. The current era of ``noisy intermediate-scale quantum'' (NISQ) devices is characterized by a relatively low number of available qubits and high physical error rates making their practical uses very limited. To overcome the problem of noise making qubits decohere before any useful calculation can be made on them, we need various techniques of quantum error correction which essentially combine physical qubits into logical units (logical qubits) at the cost of a significant overhead. The future era in which noise has been sufficiently reduced is called the era of fault tolerance (FT). It is also customary to distinguish the transition period of early fault tolerance (EFT) in which essentially fault tolerant techniques are adapted to the limitations of currently available hardware. While NISQ techniques would be more familiar in the chemistry community, in this tutorial, we will take a more future oriented stand and demonstrate the uses of fault tolerant quantum computing. Our purpose here is to take the H$_2$ example that is possible to do on current quantum hardware\cite{graham_2022,yamamoto2025quantum} and explain how it is done up to the point where error correction techniques are introduced which we will only treat here very briefly, see Section~\ref{SSec:QEC}. These are complicated techniques that deserve a treatment on their own right and have been discussed in dedicated publications in the literature. We hope that this approach will provide an educational introduction to much more complicated calculations, e.g., those on FeMoco,\cite{Reiher_2017,low2025fast} in the same vein as a detailed description of the H$_2$ case will help understand how classical approaches work. We also hope that this will fill in some of the gaps on how to obtain a Hamiltonian that the quantum computer can work with as these details are often ignored in publications focused on hardware experiments.

\subsection{The Second-Quantized Hamiltonian}
\label{SSec:2QH}

Second quantization is a technique in which the evaluation of matrix elements is performed through algebraic operations. To achieve this one switches from the Hilbert-space representation to a Fock-space representation. Within the Fock space, these Slater determinants are represented as occupation number vectors (ONV), i.e., the list of the occupation numbers of orbitals in their canonical order. Next, fermionic creation and annihilation operators are defined that map ONVs onto other ONVs.  If $n_P=0,1$ is the occupation number of spin-orbital $P$, which may be labelled using integers $P=0,1,2,\ldots$, then the annihilation and creation operators are defined by their action as
\begin{align}
    \hat{a}_P^\dagger | n_0, \ldots, n_P, \ldots, n_N \rangle & = \delta_{0n_P}\Gamma_P | n_0, \ldots, 1_P, \ldots, n_N \rangle,
    \label{annihilator_ONV}
    \\
    \hat{a}_P | n_0, \ldots, n_P, \ldots, n_N \rangle & = \delta_{1n_P}\Gamma_P | n_0, \ldots, 0_P, \ldots, n_N \rangle, 
    \label{creator_ONV}
\end{align}
where $\Gamma_P=\prod_{Q=0}^{Q=P-1}(-1)^{n_Q}$ is just a sign factor. These operators obey the following anti-commutation relations,
\begin{align}
    \{\hat{a}_{P},\hat{a}_{Q}\} &= 0,  \label{eq:fermionic_anticommutation_start} \\
    \{\hat{a}_{P}^\dagger, \hat{a}_{Q}^\dagger\} &= 0, \\
    \{\hat{a}_{P}^\dagger, \hat{a}_{Q}\} &= \delta_{PQ},
    \label{eq:fermionic_anticommutation_end}
\end{align}
and it is worth pointing out that $\delta_{PQ}=\delta_{pq}\delta_{\sigma_p\sigma_q}$, see discussion on Eq.~\eqref{spin-orb}. Then, the Hamiltonian can be written in terms of creation and annihilation operators, which is its second-quantized form, 
\begin{equation}
    \hat{H} = 
    E_n
    + \sum_{PQ}^{N_b} (P|\hat{h}|Q)
    \hat{a}_{P}^\dagger \hat{a}_{Q}
    + \frac{1}{2} \sum_{PQRS}^{N_b}  (PS|RQ) 
    \hat{a}_{P}^\dagger\hat{a}_{Q}^\dagger \hat{a}_{R}\hat{a}_{S},
    \label{Ham_SQ_ladder}
\end{equation}
where the connection with the MO integrals defined previously is $(P|\hat{h}|Q)=(p|\hat{h}|q)\delta_{\sigma_p\sigma_q}$ and $(PS|RQ)=(ps|rq)\delta_{\sigma_p\sigma_s}\delta_{\sigma_r\sigma_q}$. Note that this form of the Hamiltonian does not refer to the number of electrons as is the case in Eq.~\eqref{Ham}, but sums over all spin-orbitals, the total number of which is identical to the number of basis functions, $N_b$. It should also be mentioned that the second quantized Hamiltonian is not the only option for quantum computing although it was initially more commonly used in this field. Indeed, recently first quantized approaches\cite{su2021fault,georges2025quantum} have been studied and are a promising alternative. However, for the purposes of this tutorial, second quantization will be sufficient.

We next again consider the H$_2$ example in particular. Using the minimal basis, each of the $2$ electrons can be in $4$ possible states, the canonical order of which is $\varphi_{i}\alpha, \varphi_{i}\beta, \varphi_{a}\alpha, \varphi_{a}\beta$. Relabeling these as $P = 0,1,2,3$, a possible two-electron state has the form $|n_0, n_1, n_2, n_3 \rangle$ (with the sum of occupation numbers, i.e., the number of electrons, being 2). Due to the Pauli exclusion principle, each occupation number can be equal to $0$ or $1$. Thus, the lowest-energy determinant (ONV) is simply $|1100\rangle$, and other determinants can be written similarly. The annihilation and creation operators defined in Eq.~\eqref{annihilator_ONV} and \eqref{creator_ONV} are nothing but convenient means of converting ONVs into each other. Thus, the determinant $|1001\rangle$ can be easily obtained as
\begin{equation}
    a_3^\dagger a_1 |1100\rangle = |1001\rangle,
    \label{SQex}
\end{equation}
by the action of the operator $a_3^\dagger a_1$ removing an electron from position 1 and placing one to position 3. Setting up the FCI problem would correspond to evaluating matrix elements of $\hat{H}$ as defined in Eq.~\eqref{Ham_SQ_ladder} with respect to these Slater determinants (ONVs). As one example, calculating $\langle 1100 |\hat{H}|1100\rangle$ would yield the result in Eq.~\eqref{E0RHF}. We can also write the full H$_2$ Hamiltonian in second-quantized form. In the H$_2$ case, some of the one- and two-body integrals vanish due to spin-integration. Other terms are zero due to spatial symmetry. After simplifications, the H$_2$ Hamiltonian takes the form\cite{whitfield_simulation_2011}
\begin{align}
    \hat{H} &= 
    E_n
    + (i|\hat{h}|i) \left(a_0^\dagger a_0 + a_1^\dagger a_1\right)
    + (a|\hat{h}|a) \left(a_2^\dagger a_2 + a_3^\dagger a_3\right) \nonumber \\ 
    &+ (ii|ii)\, a_0^\dagger a_1^\dagger a_1 a_0
    +\overline{(ii|aa)} \left(a_0^\dagger a_2^\dagger a_2 a_0 + a_1^\dagger a_3^\dagger a_3 a_1\right) \nonumber \\
    &+(ii|aa) \left(a_0^\dagger a_3^\dagger a_3 a_0 + a_1^\dagger a_2^\dagger a_2 a_1\right)
    + (aa|aa)\, a_2^\dagger a_3^\dagger a_3 a_2
    \nonumber \\
    &+ (ia|ia) \left( a_0^\dagger a_3^\dagger a_1 a_2 + a_2^\dagger a_1^\dagger a_3 a_0 + a_0^\dagger a_1^\dagger a_3 a_2 + a_2^\dagger a_3^\dagger a_1 a_0 \right),
    \label{Ham_SQ}
\end{align}
where the antisymetrized integral is defined as $\overline{(ii|aa)} = (ii|aa) - (ia|ia)$. All of these integrals are known from the classical calculations in the previous section.

\subsection{Second-Quantized Qubit Mappings}
\label{SSec:SQM}

As discussed before, the basic operations on a quantum computer are carried out on two-state quantum systems called qubits. Notice that, because each spin orbital in a chemical system can be in state $|0\rangle$ or $|1\rangle$, it seems reasonable that we can map Slater determinants, and fermionic Hamiltonians, to a qubit representation. However, the operators that act on qubits are written in terms of Pauli matrices (defined in Supporting Information), which follow a different algebra compared to the fermionic creation and annihilation operators. Therefore,  we need a way to convert the fermionic operators in the second-quantized Hamiltonian to the Pauli representation. The oldest and simplest mapping is due to Jordan and Wigner,\cite{jordan1928ueber,whitfield_simulation_2011}
\begin{align}
    a_P^\dagger & \to \frac{1}{2} \left(X_P - iY_P \right) \bigotimes_{Q<P} Z_Q, 
    \label{JW1}
    \\
    a_P & \to \frac{1}{2} \left(X_P + iY_P\right) \bigotimes_{Q<P} Z_Q,
    \label{JW2}
\end{align}
where the sub-index indicates the qubit the matrix is acting on. Here, the combination of $X$ and $Y$ operators will produce $|1\rangle$ from $|0\rangle$, $|0\rangle$ from $|1\rangle$ or zero at site $P$ in a similar way as the second quantized operators change occupation numbers, while the string of $Z$ operators is needed to enforce the fermionic anti-commutation relations defined in Eqs.~(\ref{eq:fermionic_anticommutation_start}) to (\ref{eq:fermionic_anticommutation_end}). To see this, it is instructive to apply Eq.~\eqref{JW1} and \eqref{JW2} to find the Pauli string corresponding to $a_3^\dagger a_1$ in Eq.~\eqref{SQex},
\begin{align}
    a_3^\dagger a_1 
    &\to
    \frac{1}{4} \left(X_3 - iY_3 \right) Z_2 Z_1\left(X_1 + iY_1\right) Z_0^2 \nonumber \\
    &=
    \frac{1}{4} \left(X_3 - iY_3 \right) Z_2 \left(X_1 + iY_1\right),
\end{align}
where we have used the property $iY_P=Z_P X_P$ and that because $Z_P^2=I$ the identity matrix $I$ can be omitted. This operator, when acting on the sequence $|1100\rangle$ now understood as a tensor product of spin-up and spin-down states, will produce a state equivalent to that obtained in Eq.~\eqref{SQex} by the action of $a_3^\dagger a_1$ on an ONV,
\begin{equation}
    \frac{1}{4} 
    \left(X_3 - iY_3 \right) Z_2 \left(X_1 + iY_1\right)
    |1100\rangle = |1001\rangle.
\end{equation}

More generally, applying this mapping to Eq.~\eqref{Ham_SQ_ladder} leads to the qubit Hamiltonian $\hat{H}$, the explicit form of which can be found in the Supporting Information for Hamiltonians with real coefficients. While the qubit Hamiltonian has O($N_b^4$) terms in the general case, it assumes a relatively simple form in the H$_2$ case,\cite{whitfield_simulation_2011}
\begin{align}
    \hat{H} &= H_0 + H_{ii}\,[Z_0 + Z_1] + H_{aa}\,[Z_2 + Z_3] \nonumber \\
   &+ \frac{1}{4}(ii|ii) Z_0 Z_1 
    + \frac{1}{4}\overline{(ii|aa)}\, [Z_0 Z_2 + Z_1 Z_3] \nonumber \\
   &+ \frac{1}{4}(ii|aa)\, [Z_0 Z_3 + Z_1 Z_2]
    + \frac{1}{4}(aa|aa) Z_2 Z_3 \nonumber \\
   &- \frac{1}{4}(ia|ia)\, 
   [
   X_0 X_1 Y_2 Y_3
 + Y_0 Y_1 X_2 X_3
 - X_0 Y_1 Y_2 X_3
 - Y_0 X_1 X_2 Y_3
   ],
   \label{HamJW}
\end{align}
with coefficients
\begin{equation}
    H_0 = 
    E_n + (i|\hat{h}|i) + (a|\hat{h}|a) 
    + \frac{1}{4}(ii|ii) + \frac{1}{2}\overline{\overline{(ii|aa)}} + \frac{1}{4}(aa|aa),
\end{equation}
\begin{equation}
    H_{ii} = 
    \frac{1}{2}(i|\hat{h}|i)  
    + \frac{1}{4}\left[(ii|ii) + \overline{\overline{(ii|aa)}}\right],
\end{equation}
\begin{equation}
    H_{aa} = 
    \frac{1}{2}(a|\hat{h}|a)  
    + \frac{1}{4}\left[\overline{\overline{(ii|aa)}} + (aa|aa)\right].
\end{equation}
Here, the spin-summed integral is $\overline{\overline{(ii|aa)}} = 2(ii|aa)-(ia|ia)$.

There have been several alternative proposals to improve on the Jordan-Wigner mapping, both in terms of the number of qubits used and in terms of the length of the Pauli strings. A simple improvement to reduce the number of qubits required is the Qubit Efficient Encoding (QEE)\cite{shee_qubit-efficient_2022}. In this case, the mapping focuses on the fermionic ladder operators $\hat{a}^\dagger_P \hat{a}_Q$ which corresponds to sums of diadic products of basis vectors of the type $|\mathbf{n}\rangle\langle \mathbf{n}'|$, where $\mathbf{n}$ and $\mathbf{n}'$ are sequences of occupation numbers that differ at positions $P$ and $Q$ ($n_P=n'_Q=1$, $n_Q=n'_P=0$). To proceed further, the basis vectors are converted into a binary form based on their ordering. In the general case, the number of qubits required is only logarithmic in the number of spin orbitals. In the H$_2$ case, there are six possible two-electron basis states of the type $|\mathbf{n}\rangle =|n_0,n_1,n_2,n_3\rangle$ such that the occupation numbers add up to 2. In the occupation number representation, encoding $|\mathbf{n}\rangle$ requires 4 qubits. However, the six possible basis states may also be labeled as $0,\ldots,5$ by some convention. Since the binary representation of the largest ordinal, 5, is 101 and this requires only three digits, all six states can also be represented as $|\mathbf{b}\rangle = |b_0,b_1,b_2\rangle$ using only 3 qubits. The fermionic ladder operators then also have the form $|\mathbf{b}\rangle\langle \mathbf{b}'|$ which can be decomposed into tensor products of $|0\rangle\langle 0|$, $|0\rangle\langle 1|$, $|1\rangle\langle 0|$ and $|1\rangle\langle 1|$.

A separate issue with the Jordan-Wigner encoding is the long string of anti-symmetrizing $Z$ operators that appears after the mapping, which leads to undesirable scaling. Bravyi and Kitaev proposed a new mapping which encoded this anti-symmetrization in a more efficient way\cite{bravyi_2002}. The number of qubits required still depends on the number of spin orbitals, $N$, but this time, the information that is stored on these qubits depends on the qubit index, starting from 0. If the index is even then the qubit is encoded with the orbital occupation, much like in Jordan-Wigner. If the index is odd then the anti-symmetrization of a subset of orbitals is encoded. Finally, when $\log_2(i+1)$ is an integer (where $i$ is the qubit index) then the anti-symmetrization of all orbitals with an index lower than or equal to the current index is encoded. All sums are performed in modulo 2. The Jordan-Wigner and Bravyi-Kitaev mapping have been compared in the literature for chemical calculations.\cite{tranter_comparison_2018} Although the Bravyi-Kitaev approach certainly has its advantages, for our purposes, the Jordan-Wigner mapping is a sufficient starting point.

\subsection{The 1-Qubit Hydrogen Hamiltonian}

The symmetries present in the Hamiltonian can be exploited to reduce (or ``taper'') the number of qubits required for a calculation. For the case of H$_2$, note that only two Slater determinants, $|1100\rangle$ and $|0011\rangle$, can contribute to the ground-state wavefunction, due to particle number, spin and spatial symmetries. Since only two states can contribute, this suggests that the corresponding Hamiltonian can be represented by just a single qubit.

The general procedure to reduce the Hamiltonian is beyond the scope of this paper and is discussed elsewhere.\cite{bravyi_tapering_2017,setia_reducing_2020} Here, it is enough to note that we are looking for a transformation of the type
\begin{equation}
    \hat{H}'=U^\dagger \hat{H} U,
\end{equation}
where $U$ is unitary. Since $\hat{H}$ and $\hat{H}'$ are unitarily equivalent, their eigenvalues are also the same. The main requirement that should make this transformation worthwhile is that $\hat{H}'$ should commute with Pauli $X$ matrices for at least some of the qubits. If this holds, then for the purposes of determining the ground-state energy, these qubits can be replaced by the eigenvalues of the corresponding $X$ matrices, i.e., either $+1$ or $-1$. In the H$_2$ case, $U$ can be written\cite{bravyi_tapering_2017} as $U=U_1 U_2 U_3$, with
\begin{equation}
    U_P = \frac{1}{\sqrt{2}}(X_P + Z_0 Z_P),
\end{equation}
for $P = 1, 2, 3$. The transformation $\mathcal{P}'=U^\dagger\mathcal{P}U$ can now be performed for each Pauli string $\mathcal{P}$ in the Jordan-Wigner qubit Hamiltonian in Eq.~\eqref{HamJW}. The results are summarized in the Supporting Information. As a consequence, only $X$ or $I$ matrices act on qubits $1$, $2$ and $3$ in $H'$. For example, for $\mathcal{P} = Z_2$, $\mathcal{P}' = Z_0 X_2$; although there is a $Z$ acting on qubit 0, only $X$ or $I$ Paulis act on qubits $1$, $2$ and $3$.

Before these qubits can be tapered, the corresponding eigenvalues of $X$ matrices should be known for the eigenstate of $\hat{H}'$ that we seek, which will be the ground state. Here, symmetry can again be exploited since, as noted above, it only allows the configurations $\left|1100\right>$ and $\left|0011\right>$ to contribute to the ground state, $|\Psi\rangle$. Therefore, the eigenvalues of the operators $Z_0 Z_1$ must be equal to $+1$, while the eigenvalue of $Z_0 Z_2$ and $Z_0 Z_3$, will equal $-1$. Taking $Z_0 Z_1$ as an example, we may write
\begin{equation}
    Z_0 Z_1 |\Psi \rangle = | \Psi \rangle.
\end{equation}
Inserting $U U^{\dagger} = I$,
\begin{equation}
    Z_0 Z_1 U U^{\dagger} | \Psi \rangle = | \Psi \rangle,
\end{equation}
and applying $U^\dagger$ to both sides gives
\begin{equation}
    \Big( U^{\dagger} Z_0Z_1 U \Big) U^{\dagger} |\Psi \rangle = U^{\dagger} | \Psi \rangle.
\end{equation}
From the above, $U^{\dagger} | \Psi \rangle$ is an eigenvector of the transformed Hamiltonian $\hat{H}'$ and based on the results shown in the Supporting Information, $U^{\dagger} Z_0 Z_1 U = X_1$. Therefore, the eigenstates of $H'$ are also eigenstates of $X_1$ with eigenvalue $+1$, and all instances of $X_1$ in the Hamiltonian can be replaced by $+1$. The same argument can be worked through for $X_2$ and $X_3$, which will be replaced by eigenvalues $-1$.

Thus, the only operators remaining in the transformed Hamiltonian are $Z_0$ and $X_0$ acting on qubit $0$, and qubits $1$, $2$ and $3$ can be removed. The final single-qubit Hamiltonian has the form
\begin{equation}
    \hat{H}' = c_0 + c_1 Z_0 + c_2 X_0,
    \label{eq:h2_final}
\end{equation}
with
\begin{equation}
    c_0 = H_0 + \frac{1}{4}[(ii|ii) + (aa|aa)] - \frac{1}{2} \overline{\overline{(ii|aa)}},
\end{equation}
\begin{equation}
    c_1 = 2(H_{ii} - H_{aa}),
\end{equation}
\begin{equation}
    c_2 = (ia|ia).
\end{equation}
This simple Hamiltonian is ideal for minimal tests on current quantum devices and has been used for this purpose in recent studies \cite{graham_2022, blunt_2023}. We emphasize that other chemical Hamiltonians beyond H$_2$ in a minimal basis cannot be reduced to a single-qubit Hamiltonian; in general, many qubits will be needed. For a chemical system with $N$ spin orbitals a simple Jordan-Wigner transformation maps to a Hamiltonian acting on $N$ qubits. However, it is usually possible to taper at least two qubits using particle number and spin symmetries, assuming that these symmetries are present.\cite{bravyi_tapering_2017}

We will now elaborate on the most commonly used algorithms for studying such a Hamiltonian.

\subsection{Quantum Algorithms}
\label{sec:qpe}

Once the qubit Hamiltonian is available, the question still remains of how the energy calculation is to be carried out. In the current era of noisy intermediate scale quantum (NISQ) devices, the program depth measured in terms of the number of gates in the quantum circuit must be short enough so that the program can run before device errors ruin the result. This has led to a search for algorithmic solutions that satisfy this criteria, perhaps the most influential of them for chemistry being the variational quantum eigensolver (VQE) algorithm\cite{vqe_2014}. In VQE, the wavefunction is parametrized in a similar way as in traditional approaches of quantum chemistry, such as coupled cluster theory and variational Monte Carlo, except that the quantum implementation should be unitary. The use of unitary coupled cluster theory for quantum computing has been reviewed in detail elsewhere.\cite{anand2022quantum} Such approaches rely on Ans\"atze, i.e., the wavefunction is parametrized using a reference function (typically the HF solution) and parameterized quantum gates acting on it. This leads to a linear combination of excited determinants. VQE is a hybrid classical-quantum algorithm in which the energy evaluations happen on the quantum computer, while the optimization of the wavefunction coefficients is performed on the classical computer. Although this approach is more familiar to computational chemists, and, for H$_2$ in the minimal basis, it could even yield the exact energy, it has steep scaling with system size\cite{bluntPerspectiveCurrentStateoftheArt2022}, and faces challenging issues regarding optimization of the parameterized wavefunction \cite{larocca_2025}, and so we do not consider it further here.

\begin{figure}
\begin{quantikz}[row sep=0.15cm]
\lstick{$\ket{0}$} & \gate{H} & \qw & \qw      & \ \ldots\ \qw & \ctrl{4} & \gate[4, nwires=2][0.8cm]{\mathrm{QFT}^{-1}} & \meter{} & \\
\lstick{\hspace{-1cm} \vdots} & \vdots & & & \vdots & & & \vdots & \\
\lstick{$\ket{0}$} & \gate{H} & \qw & \ctrl{2} & \ \ldots\ \qw & \qw & & \meter{} & \\
\lstick{$\ket{0}$} & \gate{H} & \ctrl{1} & \qw & \ \ldots\ \qw & \qw & & \meter{} & \\
\lstick{$\ket{\psi}$} & \qw\qwbundle{n} & \gate[wires=1][0.7cm][0.7cm]{U^{2^0}} \qw & \gate[wires=1][0.7cm][0.7cm]{U^{2^1}} \qw & \ \ldots\ \qw & \gate[wires=1][0.7cm][0.7cm]{U^{2^{m-1}}} & \qw & \qw &
\end{quantikz}
\caption{The textbook quantum phase estimation (QPE) circuit. The $n$ data qubits are prepared in an initial state $| \psi \rangle$. The top $m$ qubits are ancilla qubits, which are measured at the end of the circuit to obtain an eigenphases of the unitary $U$ to $m$ bits of precision.}
\label{fig:textbook_qpe_circuit}
\end{figure}
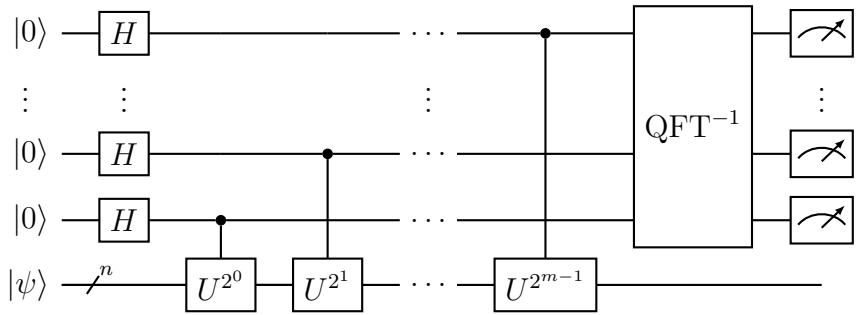

Quantum phase estimation (QPE), on the other hand, is a purely quantum algorithm, first introduced by Kitaev in 1995\cite{kitaev_quantum_1995}. The QPE method can be used to determine the eigenvalues of a unitary operator $U$,
\begin{equation}
    U |\Psi_k\rangle = e^{2\pi i \theta_k} |\Psi_k\rangle,
\end{equation}
where $\theta_k$ is the phase corresponding to the $k$'th eigenstate of $U$, $|\Psi_k\rangle$. The quantum circuit diagram for the ``textbook'' QPE algorithm\cite{nielsen_quantum_2010} is shown in Fig.~\ref{fig:textbook_qpe_circuit}. In such a circuit diagram, the individual qubits or multiqubit registers are shown as horizontal lines and their initial states are listed on the left along the y-axis, while along the x-axis the various gates acting on these qubits are listed in boxes from left to right in the order they act. The number of qubits in a circuit is often referred to as the width of the circuit or as spatial complexity (i.e., space occupied in quantum memory), while the number of gates is called circuit depth or time complexity (since gates act in a sequence in real time). The top $m$ qubits of  Fig.~\ref{fig:textbook_qpe_circuit} are ancilla qubits, which are measured at the end of the circuit to obtain the first $m$ bits of an eigenphase $\theta_k$ of $U$. The bottom $n$ qubits (represented in this circuit diagram by a single line), to which the unitary $U$ is applied, are prepared in an initial state, $|\psi\rangle$. Thus, the input of the QPE circuit can be written as $\ket{0}^{\otimes m}\otimes\ket{\psi}$ where the first $m$ qubits are reserved for measuring the eigenvalue of $U$ and the subsequent $n$ qubits contain our initial state. Here, $n$ is the number of qubits in the Hamiltonian, which is equal to $1$ for the H$_2$ Hamiltonian in Eq.~\ref{eq:h2_final}. The initial state $|\psi\rangle$ should be a good approximation to the exact eigenstate, $|\Psi_k\rangle$, whose eigenphase we want to estimate. The larger the overlap, the higher the probability of collapsing to the desired state and measuring the desired $\theta_k$. However, in general there is a chance that the wavefunction will collapse to an undesired $|\Psi_k\rangle$ upon measurement.

Remember that our goal is to estimate the eigenvalues of $\hat{H}$, but QPE provides the eigenphases of a unitary $U$. In order to apply QPE to the energy estimation problem, the eigenvalues of $\hat{H}$ must be encoded in the phases of $U$. Performing QPE with $U$ will then allow estimation of the desired energies. The most common encoding of $\hat{H}$ in $U$ is through the time evolution operator\footnote{Usually the time evolution operator would be $e^{-i\hat{H}t}$, but the minus sign is unimportant in QPE, as the phases can be extracted regardless. We call $U(t)$ the time evolution operator for brevity.},
\begin{equation}
    U(t) = e^{i \hat{H} t},
    \label{eq:time_evolution_op}
\end{equation}
where $t$ is a scalar parameter. If the eigenvalues of $\hat{H}$ are denoted $E_k$, then the eigenvalues of $U$ will be of the form $e^{i E_k t}$, and it is trivial to obtain the desired energy from the measured eigenphase. An alternative encoding is sometimes considered in more sophisticated implementations of QPE, which is discussed in Section~\ref{sec:qubitisation}. The implementation of the time evolution operator in Eq.~\ref{eq:time_evolution_op} will be considered in Section~\ref{sec:trotter}.

Before we proceed further, let us complete our discussion of the QPE circuit Fig.~\ref{fig:textbook_qpe_circuit}, specifically taking the case where $U = e^{i\hat{H}t}$. So far, we have defined a unitary operator corresponding to the Hamiltonian and we have an initial state, $\ket{0}^{\otimes m}\otimes\ket{\psi}$. The next step in the QPE algorithm is to act with a tensor product of Hadamard gates $H$ (not to be confused with the Hamiltonian $\hat{H}$) on the $m$ ancilla qubits. A Hadamard gate acting on a $\ket{0}$ state will produce an equal weight superposition on $\ket{0}$ and $\ket{1}$ states. Thus, acting with a tensor product of Hadamards on the ancillas will produce all possible basis states (all possible sequences of 0s and 1s) and we may label one of these as $\ket{s}$, where $s$ is an integer running between 0 and $2^m-1$. The next step is the controlled applications of the time evolution operator $U$. A controlled gate is a multi-qubit operation that only applies a given gate, in this case $U$ (and its powers), on the state register containing $\ket{\psi}$ if the control (marked by a black circle to which boxes of $U$s are connected vertically) is in the $\ket{1}$ state. Applying various integer powers of $U$ between 0 and $2^m-1$ in this controlled fashion will essentially attach a different power of the phase to each distinct basis state $\ket{s}$. This will result in a state
\begin{equation}
    \underbrace{\frac{1}{\sqrt{2^m}}\sum
    _{s=0}^{2^m-1}\ket{s}}_{H^{\otimes m}\ket{0}^{\otimes m}} \otimes \underbrace{\sum_k c_k \ket{\Psi_k}}_{\ket{\psi}}
    \longrightarrow
    \frac{1}{\sqrt{2^m}}  \sum_k \sum_{s=0}^{2^m-1} e^{iE_kts} \ket{s} \otimes c_k\ket{\Psi_k}.
\end{equation}
We have also made use of the fact here that the initial state can be written as a superposition of exact energy eigenstates, $\ket{\Psi_k}$, with amplitudes $c_k$. This first thing to notice at this stage is that the the first register of $m$ qubits involves a sum over the basis states $\ket{s}$ that matches the result of a Fourier transform. Thus, in the penultimate step of the QPE procedure, we apply an inverse Fourier transformation\cite{nielsen_quantum_2010} (QFT$^{-1}$ in Fig.~\ref{fig:textbook_qpe_circuit}) in the variable $s$ and obtain a state that we might write as $\sum_k c_k \ket{E_kt}\otimes\ket{\Psi_k}$, assuming that $E_kt$ can be exactly represented with $m$ bits\footnote{Otherwise, the output of the inverse Fourier transform will be a distribution peaked around $E_kt$. Note also that the time evolution operator will not be implemented exactly on a quantum computer, resulting in further inaccuracies of the distribution.}. This we can read out by measuring the first register in the computational basis $\ket{s}$, which collapses the output state to a branch $k$ of the superposition with probability $|c_k|^2$, leaving a state $\ket{s = E_kt}\otimes\ket{\Psi_k}$.

We may now try to relate this to some more familiar concepts from quantum mechanics. The general solution to the time-dependent Schr\"odinger equation is a wavepacket of the form $\sum_k c_k\ket{\Psi_k} e^{-iE_k t}$. Of crucial importance in spectroscopic applications and scattering theory is the Fourier pair made up of the autocorrelation function and the spectrum. The autocorrelation function in this context is just the overlap of the initial state and the time-evolved exact state, $\braket{\psi|\psi(t)} = \sum_k |c_k|^2 e^{-iE_kt}$, while the spectrum takes the form $\sigma(E) = \sum_k |c_k|^2\delta(E-E_k)$.\cite{Tannor_book} The QPE output can now be understood as the Fourier transformation of the autocorrelation function. Taking the (inverse) Fourier transform with respect to the register $s$, i.e.~time $ts$, is similar to taking the Fourier transformation of the wavepacket auto-correlation function\cite{Tannor_book} $ \braket{\psi(ts)|\psi} = \sum_k |c_k|^2 e^{iE_kts}$. It gives a distribution of strongly-peaked energies $E_k$, which consequently have high probability of being measured at the end of the circuit. They are weighted by $|c_k|^2$, which is why it is important to choose an initial state with sufficient overlap with the exact state, $\ket{\Psi_k}$, whose energy we wish to determine.

For most instances of $\hat{H}$ for chemistry problems, the operator $U=e^{i\hat{H}t}$ cannot be implemented exactly on a quantum computer, and we instead must consider approximate approaches such as Trotterization.

\subsection{Trotterization}
\label{sec:trotter}

As described in Section~\ref{sec:qpe}, we would like to perform QPE, the circuit diagram for which is shown in Fig.~\ref{fig:textbook_qpe_circuit}. We wish to encode the Hamiltonian in the unitary $U$ through time evolution, as defined in Eq.~\ref{eq:time_evolution_op}.

For the case of H$_2$ in a minimal basis, the Hamiltonian consists of two single-qubit Pauli operators and an identity contribution, as in Eq.~\ref{eq:h2_final}. We drop the constant shift $c_0$, so that
\begin{equation}
    \hat{H} = c_1 Z + c_2 X,
    \label{eq:h2_without_c0}
\end{equation}
and
\begin{equation}
    U(t) = e^{i(c_1 Z + c_2 X) t}.
\end{equation}
Therefore, we have to consider how this operator be implemented on a quantum computer.

To demonstrate Trotterization, we will consider a more general chemical Hamiltonian, which in its qubit form can be written
\begin{equation}
        \hat{H} = \sum_{j=1}^L H_j,
    \label{eq:general_h}
\end{equation}
where each $H_j$ consists of an $n$-qubit Pauli, $P_j$, and a coefficient $c_j$, so that $H_j = c_j P_j$, for example\footnote{More generally, each $H_j$ might be a linear combination of commuting Paulis\cite{Luis2022}; time-evolving a Hamiltonian of fully-commuting Pauli terms can be performed efficiently. Such $H_l$ terms are called fast-forwardable Hamiltonians.}.

A quantum computer supports a certain set of basis gates, which ultimately correspond to operations that are performed on physical qubits. General unitary operations must be constructed from this basis set. On current quantum computers, such as a superconducting quantum processor, a common set of native operations might include certain Pauli rotation gates, $R_P(\theta) = e^{-i (\theta/2) P}$, a CZ (controlled Z) gate, and measurement in the $Z$ basis. For fault-tolerant quantum computers, arbitrary rotation gates cannot be protected, and one must instead work with an alternative gate set, for example the Hadamard gate, the $Z$, $S$ and $T$ gates, and the CNOT gate. However, in both cases, complex multi-qubit operations such as $e^{i\hat{H}t}$, for a general $\hat{H}$, cannot be performed directly and must be approximated by sequences of gates from the basis set.

Perhaps the most well-known solution to approximately implement $U = e^{i\hat{H}t}$ for $\hat{H}$ as in Eq.~\ref{eq:general_h} is through Trotter product formulas, or Trotterization. The simplest of these is the first-order Trotter expansion
\begin{equation}
    e^{i\hat{H}t} \approx U_1(t) = \prod_{j=1}^L e^{i H_j t}.
\end{equation}
The second-order Trotter expansion is defined by
\begin{equation}
    e^{i\hat{H}t} \approx U_2(t) = \prod_{j=1}^L e^{i H_j t/2} \prod_{j=L}^1 e^{i H_j t/2}.
\end{equation}

The benefit of these expansions is that each term $e^{i H_j t}$ can be implemented in a fairly direct manner on a quantum computer. However, these product formula are approximate, unless all the terms in $\hat{H}$ commute with each other. In particular, the error on the $p$'th-order Trotter expansion can be rigorously bounded by\cite{Childs2021}
\begin{equation}
    \lVert U_p(t) - e^{i\hat{H}t} \rVert \le W_p \, t^{p+1},
\end{equation}
where $W_p$ is the Trotter error norm, which depends on the commutators of terms in the partitioning of $\hat{H}$, as in Eq.~\ref{eq:general_h}. The norm, $\lVert O \rVert$, is taken to be the spectral norm, also known as the operator norm, which is defined as the largest spectral value of the operator.

In order to manage this error, we split the time evolution operator into $m$ steps, each of size $t/m$:
\begin{equation}
    e^{i \sum_l H_l t} = \Big( e^{i \sum_l H_l t/m} \Big)^m.
\end{equation}
Each of the steps is then approximated by a Trotter formula, $U_p(t/m)$, which will become exact in the large-$m$ limit.

An important question is how many rotation gates of the form $e^{i H_j t/m}$ are needed to perform time evolution up to time $t$ with error $\epsilon$ (in the spectral norm), which we denote $N_{\textrm{gates}}(t,\epsilon)$. For the first-order Trotter formula the number of required gates is\cite{Lloyd1996} $N_{\textrm{gates, 1}}(t,\epsilon) = \mathcal{O}(t^2/\epsilon)$ while for the second-order formula\cite{Berry2006} $N_{\textrm{gates, 2}}(t,\epsilon) = \mathcal{O}(t^{3/2}/\sqrt{\epsilon})$. For a comparison of $N_{\mathrm{gates}}(t, \epsilon)$ for different simulation methods, see Ref.~\citenum{Childs2018}, and Ref.~\citenum{Childs2021} for a detailed presentation of Trotter error theory.

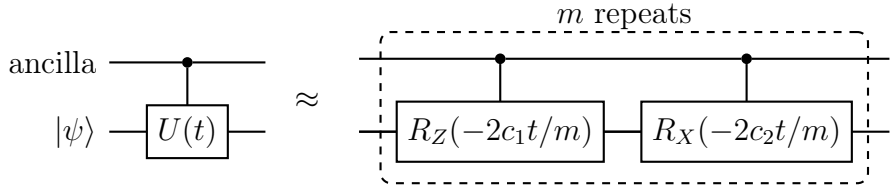
\begin{figure}
\begin{quantikz}
\lstick{$\textrm{ancilla}$} & \ctrl{1}    & \qw \\
\lstick{$\ket{\psi}$}       & \gate{U(t)} & \qw
\end{quantikz}
\; $\approx$ \;
\begin{quantikz}
\qw & \ctrl{1}\gategroup[2,steps=2,style={dashed, rounded corners, inner xsep=2pt},background]{{$m$ repetitions}} & \ctrl{1} & \qw \\
\qw & \gate[wires=1][0.8cm][0.8cm]{R_Z(-2 c_1 t/ m)} \qw & \gate[wires=1][0.8cm][0.8cm]{R_X(-2 c_2 t/ m)} \qw & \qw
\end{quantikz}
\caption{First-order Trotter approximation to the controlled time evolution operator, for H$_2$ in a minimal basis. The Trotter expansion is repeated for $m$ steps, with the approximation becoming more accurate with larger $m$.}
\label{fig:trotter_h2_circuit}
\end{figure}

Having discussed Trotterization for general Hamiltonians of the form Eq.~\ref{eq:general_h}, we now consider the specific case of H$_2$, taking the first-order Trotter expansion. Here we approximate $U(t)$ by
\begin{equation}
    U(t) \approx \Big( e^{i Z c_1 t/m} e^{i X c_2 t/m} \Big)^m,
\end{equation}
so that each term is a Pauli rotation gate. In particular, the Pauli-Z rotation is defined $R_Z(\theta) = e^{-i Z \theta/2}$, and the Pauli-X rotation is $R_X(\theta) = e^{-i X \theta/2}$. Thus we have
\begin{equation}
    U(t) \approx \biggl[ \, R_Z \biggl( \frac{- 2 c_1 t}{m} \biggr) R_X \biggl( \frac{- 2 c_2 t}{m} \biggr) \, \biggr]^m.
\end{equation}
Lastly, note from Fig.~\ref{fig:textbook_qpe_circuit} that the $U$ operators must each be controlled on an ancilla qubit. The circuit diagram for the controlled-$U$ operation is shown in Figure~\ref{fig:trotter_h2_circuit}.

We are now at the stage where we could compute the ground state energy of the H$_2$ molecule by following the QPE recipe with the specifications provided in Figure~\ref{fig:trotter_h2_circuit} if rotation gates were available on quantum devices. As we pointed out earlier, they are not and they must be broken down into other gates. The way this is done is intimately tied up with error correction techniques and is discussed in detail elsewhere.\cite{blunt_2024} We will only briefly touch upon error correction techniques in Sec.~\ref{SSec:QEC}, after we have discussed another alternative for implementing the time evolution operator in the next section.

\subsection{Qubitisation}
\label{sec:qubitisation}

Quantum phase estimation allows to measure the eigenvalues of a unitary operator $U$. Above, this was used to determine energies by choosing $U=e^{i\hat{H}t}$ and implementing the exponential in a quantum circuit using a Trotter product formula.

Alternatively, a different unitary operator $U$ can be chosen for phase estimation. In qubitisation \cite{poulinQuantumAlgorithmSpectral2018,berryImprovedTechniquesPreparing2018}, $U$ is chosen to be the walk operator
\begin{equation}
    U = e^{i\arccos (\hat{H}/\lambda)},\ \lambda = |c_1| + |c_2|
\end{equation}
with the subnormalisation $\lambda$ the norm of the Hamiltonian\footnote{To be precise, the walk operator also has eigenvalues $e^{-i\arccos(E_k/\lambda)}$.}. Performing phase estimation on the walk operator also allows to determine $\hat{H}$'s energies in a similar manner. Here, we will specifically consider the H$_2$ Hamiltonian, and the constant shift $c_0$ will again be ignored throughout.

The walk operator can be constructed from circuits called PREPARE and SELECT using an additional ancilla qubit. The ancilla qubit's state indicates the two terms $c_1Z$ and $c_2X$ of the Hamiltonian. (For larger Hamiltonians with more terms, you would require more than one ancilla qubit.)
The PREPARE operator acts on the ancilla qubit and \emph{prepares} a state corresponding to the terms' coefficients $c_1$ and $c_2$:
\begin{equation}
    \text{PREPARE}\ket{0} =  \sqrt{\frac{|c_1|}{\lambda}} \ket{0} + \sqrt{\frac{|c_2|}{\lambda}}\ket{1}= R_Y(\alpha)\ket{0},\ \alpha = 2\arctan \sqrt{\left|\frac{c_2}{c_1}\right|}.
\end{equation}
The coefficients are such that the measurement probabilities are $c_1$ and $c_2$ (here, they have the same sign, otherwise slight adaptations are necessary below), with the subnormalisation $\lambda$ being required to ensure the right-hand state is normalised. PREPARE is implemented with a single-qubit rotation gate $R_Y(\alpha)$.
The
SELECT operator acts on the system qubit and \emph{selects} the operator for the term corresponding to the state of the ancilla qubit. It applies either $Z$ or $X$ on the system qubit:
\begin{equation}
\text{SELECT}\ket{0}\ket{\psi} = \ket{0}Z\ket{\psi},\ \text{SELECT}\ket{1}\ket{\psi} = \ket{1}X\ket{\psi}.
\end{equation}
Qubitisation theory shows that the circuit for the walk operator can be constructed from these operators together with a reflection around $\ket{0}\!\bra{0}$ as follows:
\begin{equation}
\begin{tikzpicture}
\begin{yquant}
qubit {ancilla} a;
qubit {$\ket{\psi}$} psi;
box {U} (a,psi);
text {\ = \ } (-);
box {PREPARE} a;
box {SELECT} (a,psi);
box {PREPARE${}^\dagger$} a;
Z a;
\end{yquant}
\end{tikzpicture}
\end{equation}
For usage in the QPE circuit, the walk operator must be controlled on the ancillas that allow the readout of the phase. Inserting the circuits for PREPARE and SELECT in our example, we have:
\begin{equation}
\begin{tikzpicture}[baseline={([yshift=-1ex]current bounding box.center)}]
\begin{yquant}
qubit {control} c;
qubit {ancilla} a;
qubit {$\ket{\psi}$} psi;
box {U} (a,psi) | c;
text {\ = \ } (-);
box {$R_Y(\alpha)$} a;
z psi | c ~ a;
x psi | c, a;
box {$R_Y(\alpha)^\dagger$} a;
box {$Z$} a | c;
\end{yquant}
\end{tikzpicture}
\end{equation}

The advantage of this approach is that this circuit does not suffer from a Trotterisation error. Instead, it is exact (up to the finite precision of the rotation by $\alpha$). Thus, it avoids lengthy circuit repetitions stemming from a small time steps in the Trotter product formula, at the cost of an ancilla qubit. Many recent large-scale chemical quantum algorithms\cite{ivanovQuantumComputationPeriodic2023a,leeEvenMoreEfficient2021} are based on qubitisation and involve intricate circuit constructions to reduce the gate cost, and factorisations of the Hamiltonian to reduce the subnormalisation, $\lambda$.

We will now explain that the walk operator $U$ has the eigenvalues $e^{\pm i \arccos(E_k/\lambda)}$ by resorting to arguments that generalise to other and larger Hamiltonians.
Geometrically, a product of reflections about axes of relative angle $\beta$ results in a rotation by angle $2\beta$.
Both $Z$ and $\text{PREPARE}^\dagger\cdot\text{SELECT}\cdot\text{PREPARE}$ are reflections, because their squares are the identity operator. Hence, their product is a rotation. In fact, this is true individually for each eigenvalue $E_k$ of $\hat{H}$ on the two-dimensional subspace generated by $\ket{0}\ket{E_k}$. A general two-dimensional reflection matrix about an axis at inclination $\beta$ has the 
form
\begin{equation}
\begin{pmatrix}
    \cos 2\beta & -2\cos\beta\sin\beta \\
    -2\cos\beta\sin\beta & -\cos 2\beta
\end{pmatrix}.
\end{equation}
From the top left matrix element in the relevant basis generated by $\ket{0}\ket{E_k}$, we can determine the angles of the reflection axes:  $\cos 2\beta_1=\bra{0}\bra{E_k}Z\otimes I\ket{0}\ket{E_k}=+1$ for $Z$ and
\begin{equation}
\cos(2\beta_2) =\bra{0}\bra{E_k}\text{PREPARE}^\dagger\cdot\text{SELECT}\cdot\text{PREPARE}\ket{0}\ket{E_k} = E_k/\lambda.
\end{equation}
Hence, the walk operator in this basis is a rotation by angle
\begin{equation}
    \beta_\text{rot} = 2(\beta_2-\beta_1) =  \arccos(E_k/\lambda)-\arccos(+1) = \arccos(E_k/\lambda).
\end{equation}
Since a rotation by angle $\beta_\text{rot}$ has eigenvalues $e^{\pm i \beta_\text{rot}}$, the walk operator has the eigenvalues $e^{\pm  i\arccos(E_k/\lambda)}$ for each energy $E_k$ of the Hamiltonian. Having now defined an alternative way of implementing $U$, it only remains to discuss how such circuits can be implemented using the techniques of quantum error correction.

\subsection{Quantum Error Correction}
\label{SSec:QEC}

Quantum computers are affected by noise. For example, Google's latest superconducting quantum processor, Willow, which was used to perform the state-of-the-art results in Ref.~\citenum{google_2025}, reports a mean error rate of $0.33\% \pm 0.18\%$ for CZ gates \cite{willow} (noise varies strongly for different qubits and gate types, but this will suffice for the following back-of-the-envelope estimation). Roughly, this means that a quantum algorithm run on such a device could only use circuits with a depth of a few hundred operations before the error rate becomes larger than $50\%$. However, this is far too little for any useful quantum algorithm.
While the error rates of qubits are expected to decrease as technology progresses, they will always stay significant compared to error rates in classical computing. This is because qubits are inherently noisy quantum systems and even a tiny perturbation from the environment can have a disastrous effect on the qubit's state.

Luckily, quantum error correction provides a pathway to run useful longer circuits, despite the errors affecting the qubits.
To explain the concept of error correction, let us resort to the common experience of a noisy telephone line. When spelling out a name, the letters b and p can easily be confused. The error can be corrected by referring to each letter by a longer name according to the standard phonetic alphabet, like bravo for b, papa for p. This reduces the possibility of error, while increasing the length of the information transmitted.
Similarly, in quantum error correction, multiple physical qubits are used to represent one logical qubit, which has a reduced error rate compared to the physical qubit.

In quantum computing, once a qubit is measured, the wavefunction collapses and the state is destroyed. This makes it difficult to correct errors that occur in the midst of a computation. However, a theory of quantum error correction has been developed and shows intricate methods to perform measurements that reveal information about errors (if any) that have occurred, without destroying the information that is encoded. This information is sufficient to correct the errors, provided there are not too many.
For quantum algorithms of any reasonable length, we will have to resort to quantum error correction\cite{bluntPerspectiveCurrentStateoftheArt2022}.  While this results in an overhead in the number of physical qubits (many physical qubits encode one logical qubit) and run-time, it offers a chance to escape the limited fidelity of quantum computers.

\section{Summary}

In this tutorial dedicated to the memory of Prof.~Csizmadia, we have provided a detailed discussion of quantum chemical calculations on the simplest diatomic molecule, H$_2$. Such a simple exercise is not only useful as an elucidation of the traditional methods of quantum chemistry, but it also serves as an introduction to the emerging field of quantum computing, as applied to chemistry.

After providing a high-level overview of the theoretical basis of molecular calculations which led us to the Hartree--Fock model and to the notion of electron correlation, we turned to the evaluation of the necessary equations in the minimal basis. After a general discussion of the long-distance properties of the possible states in the minimal basis, the necessary integral calculations were outlined and the orbital exponent was determined by standard methods. Next, we have compared the spin-restricted Hartree--Fock and the exact solutions to discover that the exact solution removes the artifacts of the Hartree--Fock model and finds the proper covalent ground state. As a final contribution to our description of traditional methods, we have also discussed the effects of breaking spin-symmetry in the Hartree--Fock model. Next, we turned our attention to quantum computing and gave a brief discussion of second quantization in order to rewrite the Hamiltonian in terms of fermionic operators for the H$_2$ problem. We then used the Jordan-Wigner mapping to recast this Hamiltonian as a sum of Pauli-strings (products of Pauli spin-matrices) which can be implemented on a quantum computer. We have also made use of spatial symmetry to reduce the Hamiltonian to a form that acts on a single qubit. A discussion of quantum algorithms for chemistry followed, focussing on variants of quantum phase estimation. Trotterization and qubitization were introduced as two distinct algorithms for translating the single-qubit Hamiltonian into phase estimation circuits that can in principle be run on current quantum hardware.

In the final section on quantum error correction we discussed why such calculations cannot be expected to yield accurate results without applying methods to reduce noise on quantum hardware. Quantum error correction is a very active field of research which is rapidly developing, and which aims to address this problem. We note that the same H$_2$ minimal example presented in this paper was also considered in Ref.~\citenum{blunt_2024}, and compiled down to to QEC primitives, demonstrating many additional complexities in performing such a calculation in a fault-tolerant manner. Similar calculations have also been carried out on hardware recently.\cite{yamamoto2025quantum} As the field develops, we anticipate further improvements both to quantum algorithms and quantum error correction protocols, enabling quantum computers to become a powerful tool in performing practical quantum chemistry calculations.

\providecommand{\latin}[1]{#1}
\makeatletter
\providecommand{\doi}
  {\begingroup\let\do\@makeother\dospecials
  \catcode`\{=1 \catcode`\}=2 \doi@aux}
\providecommand{\doi@aux}[1]{\endgroup\texttt{#1}}
\makeatother
\providecommand*\mcitethebibliography{\thebibliography}
\csname @ifundefined\endcsname{endmcitethebibliography}
  {\let\endmcitethebibliography\endthebibliography}{}

\newpage
\begin{center}
\huge{\textbf{The Electronic Structure of the Hydrogen Molecule: A Tutorial Exercise in Classical and Quantum Computation \\ \emph{\textrm{Supporting Information}}}}
\end{center}

\setcounter{section}{0}
\setcounter{equation}{0}
\renewcommand*{\thesection}{S\arabic{section}}
\renewcommand*{\theequation}{S\arabic{equation}}

\section{Formulae for the Energy Curves and Integrals}

Molecular integrals are classified according to the operators they contain, the number of electrons they describe and the number of indices they contain ($n$-electron operators may contain up to $2n$ different indices). Hydrogen in the minimal basis will give rise to all one and two electron integrals of all the relevant operators up to two of the maximally possible four distinct indices. For one electron integrals, we will refer to the one-center integrals as atomic integrals (since all relevant quantities are centered on a single nucleus). For the overlap, these integrals give the square of the norm of the orbital. The nuclear-electron attraction integrals are special because the Coulomb potential in them also contains the coordinate of a single nucleus, and so they can run up to three different indices in general. In hydrogen, the two-index case in which the two orbitals are centered on the same atom and the Coulomb potential on the other is simply referred to the Coulomb integral, whereas all two-index one electron integrals (overlap, kinetic, attraction) in which the orbitals are centered on different atoms are called resonance orbitals. Among repulsion integrals, we again have atomic (one-center) orbitals and also hybrid integrals in which three of the orbitals share a common center. The two remaining possibilities are Coulomb integrals $(\mu\mu|\nu\nu)$ and exchange integrals $(\mu\nu|\mu\nu)$. In the following, we will summarize the values of these integrals for Gaussian orbitals and derive the corresponding energy expressions from them. The integrals themselves will be evaluated in a later section for both Gaussian and Slater orbitals.

The overlap integrals are simply given as $S_{\mu\mu}=S_{\nu\nu}=1$ and $S_{\mu\nu}=S_{\nu\mu}=e^{-2\alpha R^2}$. The kinetic energy of the electrons $T_{\mu\nu}$ is calculated as
\begin{equation}
T_{\mu\mu} = T_{\nu\nu} = -\frac{1}{2}\left(\frac{2\alpha}{\pi}\right)^{\frac{3}{2}}
\int e^{-\alpha(\mathbf{r}\pm\mathbf{R})^2}\nabla^2 e^{-\alpha(\mathbf{r}\pm\mathbf{R})^2}\,\text{d}\mathbf{r}=\frac{3}{2}\alpha,
\end{equation}
\begin{equation}
T_{\mu\nu} = T_{\nu\mu} = -\frac{1}{2}\left(\frac{2\alpha}{\pi}\right)^{\frac{3}{2}}
\int e^{-\alpha(\mathbf{r}\pm\mathbf{R})^2}\nabla^2 e^{-\alpha(\mathbf{r}\mp\mathbf{R})^2}\,\text{d}\mathbf{r}=\left(\frac{3}{2}\alpha - 2\alpha^2R^2\right)e^{-2\alpha R^2}.
\end{equation}
The nuclear-electronic attraction term also depends on the position of the nuclei. Let $A$ be the atom on which $\chi_{\mu}$ is centered and $B$ the center of $\chi_{\nu}$. Then, the total potential has the form
\begin{equation}
V_{\mu\nu} = V_{\mu\nu}(A) + V_{\mu\nu}(B),
\end{equation}
where the unique contributions are
\begin{equation}
V_{\mu\mu}(A) = V_{\nu\nu}(B) = -\left(\frac{2\alpha}{\pi}\right)^{\frac{3}{2}}
\int \frac{e^{-2\alpha(\mathbf{r}\pm\mathbf{R})^2}}{|\mathbf{r}\pm\mathbf{R}|}\,\text{d}\mathbf{r}=-2\sqrt{\frac{2\alpha}{\pi}},
\end{equation}
\begin{equation}
V_{\mu\mu}(B) = V_{\nu\nu}(A) = -\left(\frac{2\alpha}{\pi}\right)^{\frac{3}{2}}
\int \frac{e^{-2\alpha(\mathbf{r}\pm\mathbf{R})^2}}{|\mathbf{r}\mp\mathbf{R}|}\,\text{d}\mathbf{r}=-\frac{\text{erf}(2\sqrt{2\alpha}R)}{2R},
\end{equation}
\begin{equation}
V_{\mu\nu}(A) = V_{\mu\nu}(B) = -\left(\frac{2\alpha}{\pi}\right)^{\frac{3}{2}} e^{-2\alpha R^2}
\int \frac{e^{-2\alpha \mathbf{r}^2}}{|\mathbf{r}\pm\mathbf{R}|}\,\text{d}\mathbf{r}=-\frac{\text{erf}(\sqrt{2\alpha}R)}{R}e^{-2\alpha R^2}.
\end{equation}

The two-body terms can be dealt with similarly.  Since there are only two basis functions, there are only four unique integrals,
\begin{equation}
(\mu\mu|\mu\mu) = (\nu\nu|\nu\nu) = \left(\frac{2\alpha}{\pi}\right)^3 \iint \frac{e^{-2\alpha(\mathbf{r}_1\pm\mathbf{R})^2}e^{-2\alpha(\mathbf{r}_2\pm\mathbf{R})^2}}{|\mathbf{r}_1 - \mathbf{r}_2|} \,\text{d}\mathbf{r}_1\text{d}\mathbf{r}_2 = 2\sqrt{\frac{\alpha}{\pi}},
\end{equation}
\begin{equation}
(\mu\mu|\mu\nu) = (\nu\nu|\mu\nu) = \left(\frac{2\alpha}{\pi}\right)^3 e^{-2\alpha R^2} \iint \frac{e^{-2\alpha(\mathbf{r}_1\pm\mathbf{R})^2}e^{-2\alpha\mathbf{r}_2^2}}{|\mathbf{r}_1 - \mathbf{r}_2|} \,\text{d}\mathbf{r}_1\text{d}\mathbf{r}_2 = \frac{\text{erf}(\sqrt{\alpha}R)}{R}e^{-2\alpha R^2},
\end{equation}
\begin{equation}
(\mu\mu|\nu\nu) = (\nu\nu|\mu\mu) = \left(\frac{2\alpha}{\pi}\right)^3 \iint \frac{e^{-2\alpha(\mathbf{r}_1\pm\mathbf{R})^2}e^{-2\alpha(\mathbf{r}_2\mp\mathbf{R})^2}}{|\mathbf{r}_1 - \mathbf{r}_2|} \,\text{d}\mathbf{r}_1\text{d}\mathbf{r}_2 = \frac{\text{erf}(2\sqrt{\alpha}R)}{2R},
\end{equation}
\begin{equation}
(\mu\nu|\mu\nu) = (\nu\mu|\nu\mu) = \left(\frac{2\alpha}{\pi}\right)^3 e^{-4\alpha R^2} \iint \frac{e^{-2\alpha \mathbf{r}_1^2}e^{-2\alpha\mathbf{r}_2^2}}{|\mathbf{r}_1 - \mathbf{r}_2|} \,\text{d}\mathbf{r}_1\text{d}\mathbf{r}_2 = 2\sqrt{\frac{\alpha}{\pi}}e^{-4\alpha R^2}.
\end{equation}

Once the two-electron integrals are known, the most general form of the effective one-body term for two atomic orbitals can be written as
\begin{equation}
G_{\mu\mu}(\mathbf{P})
= \frac{1}{2} (\mu\mu|\mu\mu) P_{\mu\mu}
+ (\mu\mu|\mu\nu) P_{\mu\nu}
+ \frac{1}{2} (\mu\mu|\nu\nu) P_{\nu\nu},
\label{G1}
\end{equation}
\begin{equation}
G_{\nu\nu}(\mathbf{P})
= \frac{1}{2} (\nu\nu|\nu\nu) P_{\nu\nu}
+ (\nu\nu|\nu\mu) P_{\mu\nu}
+ \frac{1}{2} (\nu\nu|\mu\mu) P_{\mu\mu},
\label{G2}
\end{equation}
\begin{equation}
G_{\mu\nu}(\mathbf{P}) = G_{\nu\mu}(\mathbf{P})
= \frac{1}{2} (\mu\nu|\mu\mu) P_{\mu\mu}
+ (\mu\nu|\mu\nu) P_{\mu\nu}
+ \frac{1}{2} (\mu\nu|\nu\nu) P_{\nu\nu}.
\label{G3}
\end{equation}
Here, we have only used the fact $P_{\mu\nu}$ is symmetric and that the only molecular integral that contributes to the energy expression is $(ii|ii)$ (no exchange contribution survives). Assuming the special form the charge-density matrix $\mathbf{P}$ takes in the main text, the above equations become
\begin{equation}
G_{\mu\mu}(\mathbf{P}) = G_{\nu\nu}(\mathbf{P}) = \frac{1}{1+S_{\mu\nu}}
\left(\frac{1}{2}(\mu\mu|\mu\mu) + (\mu\mu|\mu\nu) + \frac{1}{2}(\mu\mu|\nu\nu)\right),
\end{equation}
\begin{equation}
G_{\mu\nu}(\mathbf{P}) = G_{\nu\mu}(\mathbf{P}) = \frac{1}{1+S_{\mu\nu}}
\left((\mu\mu|\mu\nu) + (\mu\nu|\mu\nu)\right).
\end{equation}
The fact that $G_{\mu\mu}=G_{\nu\nu}$ shows that the symmetry adapted orbitals are self-consistent since for a real symmetric two-by-two matrix with identical diagonal elements, the eigenvectors have the form $(a,a)$ or $(a,-a)$ for some value $a$, usually fixed by normalization. Putting all these results together, the Hartree-Fock energy for the hydrogen ground state can be calculated as
\begin{equation}
E_0 = E_{n} + \frac{1}{1+S_{\mu\nu}}(h_{\mu\mu}+h_{\mu\nu}+F_{\mu\mu}+F_{\mu\nu}),
\end{equation}
leading to
\begin{align}
E_0 &= \frac{1}{D} 
+\frac{1}{1+e^{-\frac{\alpha D^2}{2}}}
\left[
3\alpha - 4\sqrt{\frac{2\alpha}{\pi}}-\frac{2\,\text{erf}(2\sqrt{\alpha}D)}{D} +
\left(3\alpha - \alpha^2D^2 -\frac{8\,\text{erf}(\sqrt{\alpha}D)}{D}\right)e^{-\frac{\alpha D^2}{2}} 
\right.
\nonumber \\
&\left.
+\frac{1}{1+e^{-\frac{\alpha D^2}{2}}}
\left(
\sqrt{\frac{\alpha}{\pi}} +
\frac{4\,\text{erf}(\frac{\sqrt{\alpha}}{2}D)}{D}e^{-\frac{\alpha D^2}{2}} +
\frac{\text{erf}(\sqrt{\alpha}D)}{2D} +
2\sqrt{\frac{\alpha}{\pi}}e^{-\alpha D^2}
\right)
\right].
\end{align}
Here a change of variables $2R=D$ was also introduced so that the expressions depend directly on the internuclear distance $D$.

A similar process yields the following simplified $G$-elements corresponding to $\bar{\mathbf{P}}$ in the main text,
\begin{equation}
G_{\mu\mu}(\bar{\mathbf{P}}) = G_{\nu\nu}(\bar{\mathbf{P}}) = \frac{1}{1-S_{\mu\nu}}
\left(\frac{1}{2}(\mu\mu|\mu\mu) - (\mu\mu|\mu\nu) + \frac{1}{2}(\mu\mu|\nu\nu)\right),
\end{equation}
\begin{equation}
G_{\mu\nu}(\bar{\mathbf{P}}) = G_{\nu\mu}(\bar{\mathbf{P}}) = \frac{1}{1-S_{\mu\nu}}
\left((\mu\mu|\mu\nu) - (\mu\nu|\mu\nu)\right),
\end{equation}
and a new energy expression
\begin{equation}
E_1 = E_{nn} + \frac{1}{1-S_{\mu\nu}}(h_{\mu\mu}-h_{\mu\nu}+F_{\mu\mu}-F_{\mu\nu}),
\end{equation}
and finally,
\begin{align}
E_1 &= \frac{1}{D} 
+\frac{1}{1-e^{-\frac{\alpha D^2}{2}}}
\left[
3\alpha - 4\sqrt{\frac{2\alpha}{\pi}}-\frac{2\,\text{erf}(2\sqrt{\alpha}D)}{D} -
\left(3\alpha - \alpha^2D^2 -\frac{8\,\text{erf}(\sqrt{\alpha}D)}{D}\right)e^{-\frac{\alpha D^2}{2}} 
\right.
\nonumber \\
&\left.
+\frac{1}{1-e^{-\frac{\alpha D^2}{2}}}
\left(
\sqrt{\frac{\alpha}{\pi}} -
\frac{4\,\text{erf}(\frac{\sqrt{\alpha}}{2}D)}{D}e^{-\frac{\alpha D^2}{2}} +
\frac{\text{erf}(\sqrt{\alpha}D)}{2D} +
2\sqrt{\frac{\alpha}{\pi}}e^{-\alpha D^2}
\right)
\right].
\end{align}

For the singly-excited singlet state, the AO basis expression has the form
\begin{align}
E_S = \langle\Theta_S|\hat{H}|\Theta_S\rangle 
&= E_n + (P_{\mu\mu}+\bar{P}_{\mu\mu})(\mu|\hat{h}|\mu) + (P_{\mu\nu}+\bar{P}_{\mu\nu})(\mu|\hat{h}|\nu) \nonumber \\
&+ P_{\mu\mu}\bar{P}_{\mu\mu}(\mu\mu|\mu\mu) + P_{\mu\nu}\bar{P}_{\mu\nu}(\mu\nu|\mu\nu),
\end{align}
yielding
\begin{align}
E_S &= \frac{1}{D} 
+\frac{1}{1-e^{-\frac{\alpha D^2}{2}}}
\left(
3\alpha - 4\sqrt{\frac{2\alpha}{\pi}}-\frac{2\,\text{erf}(2\sqrt{\alpha}D)}{D}
\right) \nonumber \\
&-\frac{e^{-\frac{\alpha D^2}{2}}}{1-e^{-\frac{\alpha D^2}{2}}}
\left(
3\alpha - \alpha^2D^2 -\frac{8\,\text{erf}(\sqrt{\alpha}D)}{D} 
\right)
+2\sqrt{\frac{\alpha}{\pi}}.
\end{align}
Similarly for the triplets
\begin{align}
E_T = \langle\Theta_S|\hat{H}|\Theta_S\rangle 
&= E_n + (P_{\mu\mu}+\bar{P}_{\mu\mu})(\mu|\hat{h}|\mu) + (P_{\mu\nu}+\bar{P}_{\mu\nu})(\mu|\hat{h}|\nu) \nonumber \\
&+ P_{\mu\mu}\bar{P}_{\nu\nu}(\mu\mu|\nu\nu) + P_{\mu\nu}\bar{P}_{\mu\nu}(\mu\nu|\mu\nu),
\end{align}
so that
\begin{align}
E_T &= \frac{1}{D} 
+\frac{1}{1-e^{-\frac{\alpha D^2}{2}}}
\left(
3\alpha - 4\sqrt{\frac{2\alpha}{\pi}}-\frac{2\,\text{erf}(2\sqrt{\alpha}D)}{D}
\right) \nonumber \\
&-\frac{e^{-\frac{\alpha D^2}{2}}}{1-e^{-\frac{\alpha D^2}{2}}}
\left(
3\alpha - \alpha^2D^2 -\frac{8\,\text{erf}(\sqrt{\alpha}D)}{D} 
\right) \nonumber \\
&+\frac{1}{1-e^{-\frac{\alpha D^2}{2}}}
\left(
\frac{\text{erf}(\sqrt{\alpha}D)}{D}-2\sqrt{\frac{\alpha}{\pi}}e^{-\frac{\alpha D^2}{2}}
\right).
\end{align}

Finally, the off-diagonal element in the FCI-matrix in Eq.~(66) is simply given as
\begin{equation}
g = \langle\Phi_1|\hat{H}|\Phi_0\rangle = (ia|ia) = \frac{1}{1-e^{-\alpha D^2}}\left(
\sqrt{\frac{\alpha}{\pi}} - \frac{\text{erf}(\sqrt{\alpha}D)}{D}
\right).
\end{equation}

These formulae can be evaluated for any $R$ once the exponent $\alpha$ is known. We may determine this by assuming that the single Gaussian considered here is an STO-1G orbital, i.e., one in which a single Gaussian (1G) is used to fit a Slater type orbital (STO). The coefficient $\alpha$ may be obtained by maximizing the overlap
\begin{equation}
\langle \psi_{1s}|\chi_{\mu} \rangle = \sqrt{\frac{\zeta^3}{\pi}}\left(\frac{2\alpha}{\pi}\right)^{\frac{3}{4}}
\int e^{-\zeta |\mathbf{r}|}e^{-\alpha \mathbf{r}^2}\,\text{d}\mathbf{r}.
\end{equation}
Assuming that the STO exponent is $\zeta=1$, as in the H atom, this yields $\alpha\approx 0.270950$. Another way of fixing $\alpha$ is by optimizing the energy of a single H atom, $E_\text{H}$ as a function of $\alpha$. This energy is simply given as
\begin{equation}
E_{\text{H}} = T_{\mu\mu} + V_{\mu\mu}(A) = \frac{3}{2}\alpha -2\sqrt{\frac{2\alpha}{\pi}},
\end{equation}
and the optimization yields
\begin{equation}
\alpha = \frac{8}{9\pi},
\end{equation}
which is approximately $\alpha\approx 0.282942$. This choice of $\alpha$ yields the best energy value obtainable for the H atom using a single atom-centered Gaussian, $E_{\text{H}}=-\frac{4}{3}\pi\, E_\text{h}\approx -0.424413\, E_\text{h}$, still relatively far off from the exact value of $-1/2$ in atomic units. In practical calculations one would use more Gaussian functions with different exponents to get closer to the exact value. Note also that using the results of later sections, similar expressions can be found for Slater orbitals and we have provided a Mathematica file collecting all these formulae.

\section{Pauli Spin-Matrices and the Qubit Hamiltonian}

Any $2 \times 2$ matrix can be written as a linear combination of the Pauli spin-matrices $X$, $Y$, and $Z$ and the identity matrix $I$ given by
\begin{align}
    X = &
    \begin{pmatrix}
        \;0 & \;\;1\; \\
        \;1 & \;\;0\;
    \end{pmatrix}
    & Y = &
    \begin{pmatrix}
        \;0 & -i\; \\
        \;i & 0\;
    \end{pmatrix} \\
    Z = &
    \begin{pmatrix}
        \;1 & 0\; \\
        \;0 & -1\;
    \end{pmatrix}
    & I = &
    \begin{pmatrix}
        \;1 & \;\;0\; \\
        \;0 & \;\;1\;
    \end{pmatrix}
\end{align}

For a general chemical Hamiltonian with real coefficients, the explicit form of the qubit Hamiltonian in terms of MO integrals, after applying the Jordan-Wigner mapping, is
\begin{align}
\mathcal{H} & = E_n + \frac{1}{2} \left[\sum_{P}(P|\hat{h}|P) + \frac{1}{4}\sum_{PQ}\overline{(PP|QQ)}\right] \nonumber \\
& -\frac{1}{2}\sum_P \left[(P|\hat{h}|P) + \frac{1}{2}\sum_Q\overline{(PP|QQ)}\right] Z_P + \frac{1}{4}\sum_{Q<P} \overline{(PP|QQ)}\, Z_P Z_Q \nonumber \\
& +\frac{1}{2}\sum_{Q<P} \left[(P|\hat{h}|Q) + \frac{1}{4} \sum_R\overline{(PQ|RR)}\right] \left(X_P.X_Q + Y_P.Y_Q\right) \nonumber \\
& -\frac{1}{4}\sum_{P<Q<R} \overline{(PQ|RR)} Z_R\left(X_Q.X_P + Y_Q.Y_P\right) \nonumber \\
& -\frac{1}{4}\sum_{P<R<Q} \overline{(PQ|RR)} \left(X_Q.R.X_P + Y_Q.R.Y_P\right) \nonumber \\
& -\frac{1}{4}\sum_{R<P<Q} \overline{(PQ|RR)} \left(X_Q.X_P + Y_Q.Y_P\right)Z_R \nonumber \\
& -\frac{1}{4}\sum_{S<R<Q<P} [(PS|QR)-(PQ|SR)] \left(X_P.X_Q X_R.X_S + Y_P.Y_Q Y_R.Y_S\right) \nonumber \\
& -\frac{1}{4}\sum_{S<R<Q<P} [(PR|QS)-(PQ|RS)] \left(X_P.X_Q Y_R.Y_S + Y_P.Y_Q X_R.X_S\right) \nonumber \\
& -\frac{1}{4}\sum_{S<R<Q<P} [(PS|QR)-(PR|QS)] \left(X_P.Y_Q Y_R.X_S + Y_P.X_Q X_R.Y_S\right).
\end{align}
Here the notation $X_Q.X_P = X_Q Z_{Q-1}\ldots Z_{P+1} X_P$ indicates a product of $Z_S$ matrices such that $P<S<Q$, while $X_Q.R.X_P$ denotes a similar string, except that $S\neq R$.

\section{Transformations of Pauli Strings}

The transformed Pauli strings in the Jordan-Wigner Hamiltonian, after performing $\mathcal{P} \to \mathcal{P}' = U^{\dagger} \mathcal{P} U$ as defined in the main text, are as follows:
\begin{align}
    \begin{bmatrix}
        Z_0 \\
        Z_1 \\
        Z_2 \\
        Z_3 \\
        Z_0Z_1 \\
        Z_0Z_2 \\
        Z_0Z_3 \\
        Z_1Z_2 \\
        Z_1Z_3 \\
        Z_2Z_3 \\ 
        Y_0Y_1X_2X_3 \\
        X_0Y_1Y_2X_3 \\
        Y_0X_1X_2Y_3 \\
        X_0X_1Y_2Y_3 \\
    \end{bmatrix}
    \to
    \begin{bmatrix}
        Z_0 \\
        Z_0X_1 \\
        Z_0X_2 \\
        Z_0X_3 \\
        X_1 \\
        X_2 \\
        X_3 \\
        X_1X_2 \\
        X_1X_3 \\
        X_2X_3 \\ 
        X_0X_2X_3 \\
        X_0X_3 \\
        X_0X_1X_2 \\
        X_0X_1 \\
    \end{bmatrix}
\end{align}

\section{Basic Concepts for Evaluating Molecular Integrals}

\subsection{Basis Functions and Their Properties}

The simplest model of the hydrogen molecule that can be evaluated without the aid of a computer assumes that the basis functions $\chi_{\mu}(\mathbf{r})$ and $\chi_\nu(\mathbf{r})$ are simple normalized functions centred at $+\mathbf{R}$ and $-\mathbf{R}$ , with $\mathbf{R}=(0,0,R)$.  Here, we have assumed, without loss of generality, that the two nuclei are along the $z$-axis, at $z=\pm R$. Since these vectors play a special role, we will reserve the special notation for their length, $r_{\pm}=|\mathbf{r}\pm\mathbf{R}|=|(x,y,z\pm R)|$. The most common choices are Gaussian and Slater type orbitals (GTOs and STOs). Here we will consider only the simplest (s-type) GTOs and STOs.

Gaussians decompose into
\begin{equation}
\chi_{\mu}^G(\mathbf{r})
=
N_G e^{-\alpha(\mathbf{r}\pm\mathbf{R})^2}
=
N_G
e^{-\alpha x^2}
e^{-\alpha y^2}
e^{-\alpha (z\pm R)^2},
\label{Gauss}
\end{equation}
Note the change of signs: the Gaussian at $+R$ is the one with $z-R$ in the exponent, and \emph{vice versa}.  Slater type orbitals have the form
\begin{equation}
\chi_{\mu}^S(\mathbf{r})
=
N_S e^{-\alpha|\mathbf{r}\pm\mathbf{R}|}
=
N_S
e^{-\alpha \sqrt{x^2 + y^2 + (z\pm R)^2}},
\label{Slater}
\end{equation}
and are usually much harder to handle. $N_G$ and $N_S$ are the appropriate norm factors. Among other things, Gaussians are preferred for their multiplication properties. When two different Gaussians are multiplied, the Gaussian product theorem applies
\begin{equation}
e^{-\alpha(\mathbf{r}-\mathbf{A})^2}e^{-\beta(\mathbf{r}-\mathbf{B})^2} = e^{-\frac{\alpha\beta}{\alpha+\beta} (\mathbf{A}-\mathbf{B})^2} e^{-(\alpha+\beta)(\mathbf{r}-\mathbf{P})^2},
\quad
\mathbf{P}=\frac{1}{\alpha+\beta}(\alpha\mathbf{A}+\beta\mathbf{B}),
\label{GPT}
\end{equation}
where $\mathbf{P}$ is a point on the line connecting $\mathbf{A}$ and $\mathbf{B}$.
This can be appropriately generalized to more complicated Gaussians not discussed in this paper. No such simple rule exists for STOs. For the GTOs lying on the $z$-axis at equal distance from the origin, one obtains the simple form,
\begin{equation}
e^{-\alpha(\mathbf{r}\pm\mathbf{R})^2}e^{-\alpha(\mathbf{r}\mp\mathbf{R})^2} = e^{-2\alpha R^2} e^{-2\alpha\mathbf{r}^2}.
\label{GPT0}
\end{equation}
Given the relative simplicity of quantities containing GTOs,  it is sometimes convenient to convert STOs into GTOs using
\begin{equation}
e^{-\alpha r} = \frac{\alpha}{2\sqrt{\pi}}\int^{\infty}_{0} \frac{1}{\sqrt{\xi^3}} e^{-\frac{\alpha^2}{4\xi}} e^{-\xi r^2}\,\text{d}\xi.
\label{STO_to_GTO}
\end{equation}

\subsection{Integrals and Integration Techniques}

Given these choices, the calculation of electronic energies ultimately boils down to evaluating integrals of the type,
\begin{equation}
\int^\infty_{-\infty} f(\mathbf{r})\,\text{d}\mathbf{r} \equiv 
\int^\infty_{-\infty}\int^\infty_{-\infty}\int^\infty_{-\infty} f(x,y,z)\,\text{d}x\,\text{d}y\,\text{d}z,
\label{GenInt}
\end{equation}
or similar integrals with two sets of variables $\mathbf{r}_1$ and $\mathbf{r}_2$. Here $f$ is some product of GTOs and/or STOs and possibly some other terms associated with the operators.

Among standard integration techniques, integration by parts has a particularly simple form when applied to GTOs and STOs, since they and their derivatives vanish at infinity,
\begin{equation}
\int^\infty_{-\infty} \chi_\mu(\mathbf{r})\partial_i\chi_\nu(\mathbf{r})\,\text{d}\mathbf{r} =
-\int^\infty_{-\infty} \partial_i\chi_\mu(\mathbf{r})\chi_\nu(\mathbf{r})\,\text{d}\mathbf{r},
\end{equation}
where $\partial_i\chi$ denotes the partial derivative of $\chi$ with respect to $i=x,y,$ or $z$. Since $\nabla^2 = \partial^2_x + \partial^2_y + \partial^2_z$, this also conveniently yields
\begin{equation}
\int^\infty_{-\infty} \chi_\mu(\mathbf{r})\nabla^2\chi_\nu(\mathbf{r})\,\text{d}\mathbf{r} =
-\int^\infty_{-\infty} \nabla\chi_\mu(\mathbf{r})\cdot\nabla\chi_\nu(\mathbf{r})\,\text{d}\mathbf{r},
\label{IParts}
\end{equation}
with  the gradient vector being $\nabla\chi=(\partial_x\chi,\partial_y\chi,\partial_z\chi)$.
Integration by substitution comes up in various guises. For example, a displacement in an integral like Eq.~\eqref{GenInt} can be eliminated by the change of variables $\mathbf{r}'=\mathbf{r}\pm \mathbf{R}$. We will also frequently encounter the following integral
\begin{equation}
\int^\infty_{0} \frac{f\left(\frac{t^2}{a+t^2}\right)}{(a+t^2)^{\frac{3}{2}}}\,\text{d}t =
\frac{1}{a}\int^1_{0} f(u^2)\,\text{d}u,
\label{ISubs}
\end{equation}
with the substitution $u^2=t^2/(a+t^2)$, $\text{d}u=a/(a+t^2)^{\frac{3}{2}}$ and the appropriate changes in the integration limits.

Integrals can often be more conveniently evaluated in non-Cartesian coordinate systems. Among these spherical polar coordinates are perhaps the most common. These consist of the radial distance form the origin, $0\leq r$, the polar angle between the $z$-axis (the polar axis) and the radial line, $0\leq\theta\leq\pi$, and the azimuthal angle, $0\leq\varphi < 2\pi$, which is the angle of rotation around the $z$-axis. The appropriate conversions to Cartesian coordinates are
\begin{equation}
x=r\sin\theta\cos\varphi,\quad
y=r\sin\theta\sin\varphi,\quad
z=r\cos\theta.
\label{CartSpher}
\end{equation}
When it comes to integration over the full space, the appropriate volume element is converted as $\text{d}\mathbf{r}=r^2\sin\theta\text{d}r\,\text{d}\theta\,\text{d}\varphi$, 
\begin{equation}
\int^\infty_{-\infty} f(\mathbf{r})\,\text{d}\mathbf{r} = 
\int^{\infty}_{0}\int^{\pi}_{0}\int^{2\pi}_{0} r^2 \tilde{f}(r,\theta,\varphi) \sin\theta\, \text{d}\varphi\, \text{d}\theta\, \text{d}r,
\end{equation}
where $\tilde{f}$ denotes the transformed function corresponding to the function $f$ after the substitutions in Eq.~\eqref{CartSpher} have been carried out. This is particularly advantageous if the function $\tilde{f}$ only depends on $r$,
\begin{equation}
\int^\infty_{-\infty} f(\mathbf{r})\,\text{d}\mathbf{r} = 
4\pi\int^{\infty}_{0} r^2 \tilde{f}(r)\, \text{d}r.
\label{SPCnoAngle}
\end{equation}

Another useful transformation involves prolate spheroidal coordinates.  Like the spherical coordinates, these also involve rotation around the $z$-axis, but the plane rotated is mapped differently. Two points along the $z$-axis at $\pm R$ serve as two focal points for a series of ellipses and hyperbolae identified by the parameters $1\leq\sigma$ and $-1\leq\tau\leq 1$, respectively, while the azimuthal angle, $0\leq\varphi < 2\pi$, behaves similarly as in the spherical case. The conversion to Cartesian coordinates is given by
\begin{equation}
x=R\sqrt{(\sigma^2-1)(1-\tau^2)}\cos\varphi,\quad
y=R\sqrt{(\sigma^2-1)(1-\tau^2)}\sin\varphi,\quad
z=R\sigma\tau.
\end{equation}
It should be noted that the distances $r_\pm$ have an especially simple form in these coordinates, $r_\pm = R(\sigma\pm\tau)$. The appropriate volume element for integration is $\text{d}\mathbf{r} = R r_+ r_- \text{d}\sigma\,\text{d}\tau\,\text{d}\varphi = R^3 (\sigma^2-\tau^2) \text{d}\sigma\,\text{d}\tau\,\text{d}\varphi$,
\begin{equation}
\int^\infty_{-\infty} f(\mathbf{r})\,\text{d}\mathbf{r} = 
R^3 \int^{\infty}_{1}\int^{1}_{-1}\int^{2\pi}_{0} (\sigma^2-\tau^2) \tilde{f}(\sigma,\tau,\varphi) \,\text{d}\varphi\, \text{d}\tau\, \text{d}\sigma,
\end{equation}
and, if there is no angle dependence in $\tilde{f}$,
\begin{equation}
\int^\infty_{-\infty} f(\mathbf{r})\,\text{d}\mathbf{r} = 
2\pi R^3 \int^{\infty}_{1}\int^{1}_{-1} (\sigma^2 - \tau^2) \tilde{f}(\sigma,\tau)\, \text{d}\tau\, \text{d}\sigma.
\label{PECnoAngle}
\end{equation}

\subsection{The Coulomb Operator}

The integrals to be discussed here are often difficult to handle because of the presence of the Coulomb potential. One way to make it more manageable is to convert it into a Gaussian,
\begin{equation}
\frac{1}{r}=\frac{1}{\sqrt{\pi}}\int^{\infty}_{-\infty} e^{-r^2 t^2}\,\text{d}t 
= \frac{2}{\sqrt{\pi}}\int^{\infty}_{0} e^{-r^2 t^2}\,\text{d}t.
\label{Cou_to_Gau}
\end{equation}
This is especially advantageous if there are other Gaussian functions in the integral. Several other techniques exist in cases when this solution is not ideal.

One of the earliest expansions for the inverse distance between two points is due to Legendre in the 18th century. Indeed, the Legendre polynomials $P_l$ were introduced as the coefficients in the expansion
\begin{equation}
\frac{1}{|\mathbf{r}_1-\mathbf{r}_2|} = \sum_{l=0}^{\infty} \frac{r_<^l}{r_>^{l+1}} P_l (\cos\gamma),
\label{LegExp}
\end{equation}
where $r_< = \min (r_1,r_2)$, $r_> = \max (r_1,r_2)$ and $\gamma$ is the angle between $\mathbf{r}_1$ and $\mathbf{r}_2$. For our purposes, it is enough to know that the Legendre polynomials of the first kind, $P_l(x)$, can be obtained from the recursive formula,
\begin{equation}
P_l (x) = \frac{2l-1}{l}xP_{l-1}(x) - \frac{l-1}{l}P_{l-2}(x),
\end{equation}
with the first few functions being
\begin{equation}
P_0(x) = 1,\quad
P_1(x)=x,\quad
P_2(x)=\frac{1}{2}(3x^2-1).
\end{equation}
These polynomials are usually defined and used within the region $|x|\leq 1$, but the identical functions can also be used in the region $1<x$, as we will see shortly. These polynomials are special solutions of a differential equation which also admits other types of solutions.  Among these, we will be interested in Legendre polynomials of the second kind, $Q_l(x)$, with a similar recursive formula
\begin{equation}
Q_l (x) = \frac{2l-1}{l}xQ_{l-1}(x) - \frac{l-1}{l}Q_{l-2}(x),
\end{equation}
with
\begin{equation}
Q_0(x) = \frac{1}{2}\ln\frac{x+1}{|x-1|},\quad
Q_1(x)=\frac{1}{2}x\ln\frac{x+1}{|x-1|}-1,\quad
Q_2(x)=\frac{1}{4}(3x^2-1)\ln\frac{x+1}{|x-1|}-\frac{3}{2}x.
\end{equation}
The expression $|x-1|$ may be simplified depending on whether the polynomials are used in the regime $|x|\leq 1$ or $1<x$. As far as integration properties are concerned, the most important of these is the orthogonality of $P_l$ within $|x|\leq 1$
\begin{equation}
\int^{1}_{-1} P_k(x)P_l(x)\,\text{d}x = \frac{2}{2l+1}\delta_{kl}.
\label{Portho}
\end{equation}
Furthermore,
\begin{equation}
\int^{1}_{-1} P_l(x)\,\text{d}x =2\delta_{l0},
\label{Pint}
\end{equation}
which follows from Eq.~\eqref{Portho} for the case $k=0$ (since $P_0(x)=1$), and
\begin{equation}
\int^{1}_{-1} x^k P_l(x)\,\text{d}x = 0, \quad k<l.
\label{Ppow}
\end{equation}
Finally, in some of the formulas below, the associated Legendre polynomials of the first ($P^m_l$) and second kind $Q^m_l$ will also appear. These are related to the $m$th derivatives of $P_l\equiv P^0_l$ and $Q_l\equiv Q^0_l$. It will not be necessary to discuss the properties of these functions, since we will be only interested in case $m=0$. 

The expression in Eq.~\eqref{LegExp} can be rewritten so that it depends on the spherical coordinates of the two vectors,
\begin{equation}
\frac{1}{r_{12}} = \sum_{l=0}^{\infty} \frac{r_<^l}{r_>^{l+1}} \sum_{m=0}^{l} (2-\delta_{m0})\frac{(l-m)!}{(l+m)!} P_l^m (\cos\theta_1)P_l^m (\cos\theta_2)\cos m(\varphi_1-\varphi_2).
\label{SExp1}
\end{equation}
This allows for independent integration of radial and angular variables and it is especially simple for functions $f$ that, when transformed to spherical coordinates, only depend on radial variables. In that case, the integration over $\varphi_1$ and $\varphi_2$ produces a non-zero result (a factor of $4\pi^2$) if and only if $m=0$, while the integration over $\theta_1$ and $\theta_2$ is essentially the same as Eq.~\eqref{Ppow} and would only yield a non-zero factor of $4$ in the case $l=0$,
\begin{align}
&\int^{\infty}_{-\infty}\int^{\infty}_{-\infty} \frac{f(\mathbf{r}_1,\mathbf{r}_2)}{|\mathbf{r}_1-\mathbf{r}_2|}\,\text{d}\mathbf{r}_1\,\text{d}\mathbf{r}_2 = \nonumber \\
&16\pi^2\int^\infty_0\int^{r_1}_0 r_1 r_2^2\tilde{f}(r_1,r_2)\,\text{d}r_2\,\text{d}r_1 +
16\pi^2\int^\infty_0\int^\infty_{r_1} r_1^2 r_2\tilde{f}(r_1,r_2)\,\text{d}r_2\,\text{d}r_1.
\label{SExp2}
\end{align}
Since $l=0$, the only term that survives of the radial factor in Eq.~\eqref{SExp1} is $1/r_>$, which changes the integration limits depending on whether $r_1>r_2$ or $r_1<r_2$. It is also enough to calculate one of these terms if $\tilde{f}(r_1,r_2)=\tilde{f}(r_2,r_1)$.

The expansions in Eq.~\eqref{SExp1} and Eq.~\eqref{SExp2} both assume a common origin for $\mathbf{r}_1$ and $\mathbf{r}_2$, which makes them ideal for quantities where this is actually so. However, in the hydrogen molecule, there are two nuclear centers, and the question arises whether there are expansions that would facilitate the evaluation of quantities containing two vectors with two different origins, let us call them $\mathbf{r}_{1\pm}$ and $\mathbf{r}_{2\mp}$. It is possible to design such an expansion using polar coordinates,
\begin{equation}
\frac{1}{r_{12}} = \sum_{l_1=0}^{\infty} \sum_{l_2=0}^{\infty} \sum_{m=0}^{l_<} (2-\delta_{m0}) B_{l_1,l_2}^{m}(r_{1_\pm},r_{2_\mp},R) P_{l_1}^m (\cos\theta_{1_\pm})P_{l_2}^m (\cos\theta_{2_\mp})\cos m(\varphi_{1_\pm}-\varphi_{2_\mp}),
\end{equation}
where $B_{l_1,l_2}^{m}$ depends on various powers of $r_{1_\pm}$, $r_{2_\mp}$ and $R$. Since $B_{l_1,l_2}^{m}$ is also obtained from an integration in which the radial and angular variables cannot always be separated, this expression can be difficult to use. We have also seen that $r_{1_\pm}$ and $r_{2_\mp}$ have an especially simple form in prolate ellipsoidal coordinates. Thus, we turn next to an expansion due to Neumann in the 19th century that uses this coordinate system,
\begin{align}
\frac{1}{r_{12}} &= \frac{1}{R}\sum_{l=0}^{\infty} \sum_{m=0}^{l} (2-\delta_{m0})(-1)^m (2l+1) \left(\frac{(l-m)!}{(l+m)!}\right)^2 \nonumber \\
&\times P_l^m (\sigma_<)Q_l^m (\sigma_>) P_l^m (\tau_1)P_l^m (\tau_2)\cos m(\varphi_1-\varphi_2).
\label{NExp1}
\end{align}
Again, this complicated expression becomes much simpler in the case in which the a function $f(\mathbf{r}_1,\mathbf{r}_2)$ when converted to prolate ellipsoidal coordinates only depends on $\sigma_1$ and $\sigma_2$, $\tilde{f}(\sigma_1,\sigma_2)$. Integration over $\varphi_1$ and $\varphi_2$ again yields $4\pi^2$ in this case. Integration over the variables $\tau_1$ and $\tau_2$ is somewhat more tedious, but elementary,
\begin{equation}
\int^{1}_{-1}\int^{1}_{-1} (2l+1) (\sigma_1^2-\tau_1^2)(\sigma_2^2-\tau_2^2) P_l (\tau_1)P_l (\tau_2) \,\text{d}\tau_1\,\text{d}\tau_2 = \frac{16}{9}P_2(\sigma_1)P_2(\sigma_2)\delta_{l0}+\frac{16}{45}\delta_{l2}.
\label{NExp2}
\end{equation}
Most of the terms in the above integral will only survive if $l=0$ because of Eq.~\eqref{Pint}. However, because of the presence of terms containing $\tau_1^2$ and $\tau_2^2$, by Eq.~\eqref{Ppow}, it must be true that $l\leq 2$. The $l=1$ case vanishes, and thus we are left with the cases $l=0$ and $l=2$, with terms containing $\sigma_1$ and $\sigma_2$ conveniently forming polynomials $P_2$. On substituting Eq.~\eqref{NExp2} into Eq.~\eqref{NExp1},
\begin{align}
&\int^{\infty}_{-\infty}\int^{\infty}_{-\infty} \frac{f(\mathbf{r}_1,\mathbf{r}_2)}{|\mathbf{r}_1-\mathbf{r}_2|}\,\text{d}\mathbf{r}_1\,\text{d}\mathbf{r}_2 \nonumber\\
&=\frac{64\pi^2R^5}{9}\int^\infty_1\int^{\sigma_1}_1 \tilde{f}(\sigma_1,\sigma_2) Q_0 (\sigma_1)P_2(\sigma_1)P_2(\sigma_2)\,\text{d}\sigma_2\text{d}\sigma_1 \nonumber\\
&+\frac{64\pi^2R^5}{9}\int^\infty_1\int^\infty_{\sigma_1} \tilde{f}(\sigma_1,\sigma_2) Q_0 (\sigma_2)P_2(\sigma_1)P_2(\sigma_2)\,\text{d}\sigma_2\text{d}\sigma_1 \nonumber\\
&+\frac{64\pi^2R^5}{45}\int^\infty_1\int^{\sigma_1}_1 \tilde{f}(\sigma_1,\sigma_2) P_2 (\sigma_2)Q_2 (\sigma_1)\,\text{d}\sigma_2\text{d}\sigma_1 \nonumber\\
&+\frac{64\pi^2R^5}{45}\int^\infty_1\int^\infty_{\sigma_1} \tilde{f}(\sigma_1,\sigma_2) P_2 (\sigma_1)Q_2 (\sigma_2)\,\text{d}\sigma_2\text{d}\sigma_1,
\label{NeuSigma}
\end{align}
where again, the effect of $\sigma_>$ and $\sigma_<$ shows in the integration limits. For the case $\tilde{f}(\sigma_1,\sigma_2)=\tilde{f}(\sigma_2,\sigma_1)$, one of the first or the second, and one of the third and fourth terms is enough.

\subsection{List of Useful Integrals}

Some textbook results will be useful in our work. Some special functions are defined as integrals, including the error function, $\text{erf}$, and the exponential integral, $\text{Ei}$,
\begin{equation}
\text{erf}(x) = \frac{2}{\sqrt{\pi}}\int^x_0 e^{-t^2}\,\text{d}t = 
\frac{2x}{\sqrt{\pi}}\int^1_0 e^{-x^2 t^2}\,\text{d}t,
\label{IErf}
\end{equation}
\begin{equation}
\text{Ei}(x) = \int^{x}_{-\infty} \frac{e^t}{t}\,\text{d}t = -\int^{\infty}_{-x} \frac{e^{-t}}{t}\,\text{d}t.
\label{ExpI}
\end{equation}
The following definite integrals are especially useful,
\begin{equation}
\int^{\infty}_{0} x^{2n}e^{-ax^2}\,\text{d}x =
\frac{1}{2}\int^{\infty}_{-\infty} x^{2n}e^{-ax^2}\,\text{d}x =
\frac{(2n)!}{n!2^{2n+1}}\sqrt{\frac{\pi}{a^{2n+1}}},
\label{Gevn}
\end{equation}
\begin{equation}
\int^{\infty}_{0} x^{2n+1}e^{-ax^2}\,\text{d}x =
\frac{n!}{2a^{n+1}},
\label{Godd}
\end{equation}
\begin{equation}
\int^{\infty}_{0} x^{n}e^{-ax}\,\text{d}x =
\frac{n!}{a^{n+1}},
\label{Sint}
\end{equation}
\begin{equation}
\int^{\infty}_{0} e^{-x}\ln x \,\text{d}x = -\gamma,
\label{EulerGamma}
\end{equation}
\begin{equation}
\int^\infty_0 e^{-a x^2 - b/x^2}\,\text{d}x = \frac{1}{2}\sqrt{\frac{\pi}{a}} e^{2\sqrt{ab}},
\label{GauX1X}
\end{equation}
\begin{equation}
\int^\infty_0 x e^{-a x^2}\text{erf}\left(\frac{b}{x}\right)\,\text{d}x = \frac{1}{2a}(1-e^{-2b\sqrt{a}}),
\label{xErfGau}
\end{equation}
as well as the following indefinite integrals,
\begin{equation}
\int x^n e^{a x}\,\text{d}x = e^{a x} \sum_{i=0}^n (-1)^{n-i}\frac{n!}{i! a^{n-i+1}}x^i,
\label{Spoly}
\end{equation}
\begin{equation}
\int e^{ax}\ln x\,\text{d}x = \frac{1}{a} (e^{ax}\ln |x| - \text{Ei}(ax)),
\label{IExpLn}
\end{equation}
where $\gamma=0.57721...$ is Euler's constant, $a$ is a non-negative real constant and $n$ is a non-negative integer.

\section{Gaussian Integrals}

\subsection{Overlap Integrals}

\noindent\textbf{Atomic Integral (Norm).} To find the norm factor, $N_G$, we must evaluate the integrals for the product of two such functions (see Eq.~\eqref{Gauss}),
\begin{equation}
N_G^2\int^{\infty}_{-\infty} e^{-2\alpha(\mathbf{r}\pm\mathbf{R})^2}\,\text{d}\mathbf{r} =
N_G^2\int^{\infty}_{-\infty} e^{-2\alpha\mathbf{r}^2}\,\text{d}\mathbf{r} = 
N_G^2\left(\frac{\pi}{2\alpha}\right)^{\frac{3}{2}}.
\end{equation}
Because Gaussians can be factorized, the last integral is just a product of 3 Gaussian integrals with the known result in Eq.~\eqref{Gevn}. Setting this integral to 1 yields the norm factor $N_G$ in the normalized Gaussians,
\begin{equation}
\chi^G_{\mu} = \left(\frac{2\alpha}{\pi}\right)^{\frac{3}{4}}e^{-\alpha(\mathbf{r}+\mathbf{R})^2},
\quad
\chi^G_{\nu} = \left(\frac{2\alpha}{\pi}\right)^{\frac{3}{4}}e^{-\alpha(\mathbf{r}-\mathbf{R})^2}.
\end{equation}
Since the atomic overlap integrals are defined as
\begin{equation}
S_{\mu\mu} = S_{\nu\nu} = 
\left(\frac{2\alpha}{\pi}\right)^{\frac{3}{2}}
\int^{\infty}_{-\infty} e^{-2\alpha(\mathbf{r}\pm\mathbf{R})^2}\,\text{d}\mathbf{r},
\end{equation}
they are, by normalization,
\begin{equation}
\boxed{S_{\mu\mu} = S_{\nu\nu} =1.}
\end{equation}

\noindent\textbf{Resonance Integral.} Similarly, the two-centered case is simply defined as
\begin{equation}
S_{\mu\nu} = S_{\nu\mu} = 
\left(\frac{2\alpha}{\pi}\right)^{\frac{3}{2}} 
\int^{\infty}_{-\infty} e^{-\alpha (\mathbf{r}\pm \mathbf{R})^2}e^{-\alpha (\mathbf{r}\mp \mathbf{R})^2}\,\text{d}\mathbf{r},
\end{equation}
and, by Eq.~\eqref{GPT0}, it is found to be
\begin{equation}
\boxed{S_{\mu\nu} = S_{\nu\mu} = e^{-2\alpha R^2}.}
\end{equation}

\subsection{Kinetic Energy Integrals}

\noindent\textbf{Atomic Integral.} The one-center kinetic energy of the electrons $T_{\mu\mu}$ is calculated as
\begin{equation}
T_{\mu\mu} = T_{\nu\nu} = -\frac{1}{2}\left(\frac{2\alpha}{\pi}\right)^{\frac{3}{2}}
\int^{\infty}_{-\infty} e^{-\alpha(\mathbf{r}\pm\mathbf{R})^2}\nabla^2 e^{-\alpha(\mathbf{r}\pm\mathbf{R})^2}\,\text{d}\mathbf{r}.
\end{equation}
The gradient vector for the GTOs is $\nabla\chi (\mathbf{r}_{\pm}) = -2\alpha\mathbf{r}_{\pm}\chi (\mathbf{r}_{\pm})$, and by Eq.~\eqref{IParts}, we have
\begin{equation}
\int^{\infty}_{-\infty} \nabla e^{-\alpha(\mathbf{r}\pm\mathbf{R})^2}\cdot \nabla e^{-\alpha(\mathbf{r}\pm\mathbf{R})^2}\,\text{d}\mathbf{r} =
4\alpha^2 \int^{\infty}_{-\infty} r_{\pm}^2 e^{-2\alpha r_{\pm}^2} \,\text{d}\mathbf{r}=
4\alpha^2 \int^{\infty}_{-\infty} r^2 e^{-2\alpha r^2} \,\text{d}\mathbf{r},
\end{equation}
after a change of variables. Finally, switching to polar coordinates via Eq.~\eqref{SPCnoAngle} and using Eq.~\eqref{Gevn},
\begin{equation}
16\pi\alpha^2 \int^{\infty}_{0} r^4 e^{-2\alpha r^2} \,\text{d}r=
3\alpha\left(\frac{\pi}{2\alpha}\right)^{\frac{3}{2}}.
\end{equation}
Thus, we have
\begin{equation}
\boxed{T_{\mu\mu} = T_{\nu\nu} = \frac{3}{2}\alpha.}
\end{equation}

\noindent\textbf{Resonance Integral.} The remaining kinetic energy integrals have the form
\begin{equation}
T_{\mu\nu} = T_{\nu\mu} = -\frac{1}{2}\left(\frac{2\alpha}{\pi}\right)^{\frac{3}{2}}
\int^{\infty}_{-\infty} e^{-\alpha(\mathbf{r}\pm\mathbf{R})^2}\nabla^2 e^{-\alpha(\mathbf{r}\mp\mathbf{R})^2}\,\text{d}\mathbf{r}.
\end{equation}
Again, using Eq.~\eqref{IParts},
\begin{equation}
\int^{\infty}_{-\infty} \nabla e^{-\alpha(\mathbf{r}\pm\mathbf{R})^2}\cdot \nabla e^{-\alpha(\mathbf{r}\mp\mathbf{R})^2}\,\text{d}\mathbf{r} =
4\alpha^2e^{-2\alpha R^2} \int^{\infty}_{-\infty} (r^2-R^2) e^{-2\alpha r^2} \,\text{d}\mathbf{r},
\end{equation}
which in polar coordinates becomes
\begin{equation}
16\pi\alpha^2 e^{-2\alpha R^2} \int^{\infty}_{0} r^2(r^2-R^2) e^{-2\alpha r^2} \,\text{d}r
=\left(\frac{\pi}{2\alpha}\right)^{\frac{3}{2}}(3\alpha - 4\alpha^2 R^2) e^{-2\alpha R^2},
\end{equation}
which leads to
\begin{equation}
\boxed{T_{\mu\nu} = T_{\nu\mu} = \left(\frac{3}{2}\alpha - 2\alpha^2R^2\right)e^{-2\alpha R^2}.}
\end{equation}

\subsection{Nuclear-Electron Attraction Integrals}

\noindent\textbf{Atomic Integral.} The nuclear-electronic attraction term also depends on the position of the nuclei. Let $A$ be the atom on which $\chi_{\mu}$ is centered and $B$ the center of $\chi_{\nu}$. Then, the total potential has the form
\begin{equation}
V_{\mu\nu} = V_{\mu\nu}(A) + V_{\mu\nu}(B).
\end{equation}
The first unique contribution is
\begin{equation}
V_{\mu\mu}(A) = V_{\nu\nu}(B) = -\left(\frac{2\alpha}{\pi}\right)^{\frac{3}{2}}
\int^{\infty}_{-\infty} \frac{e^{-2\alpha(\mathbf{r}\pm\mathbf{R})^2}}{|\mathbf{r}\pm\mathbf{R}|}\,\text{d}\mathbf{r},
\end{equation}
and after a shift of coordinates, the integral may be converted to spherical coordinates,
\begin{equation}
\int^{\infty}_{-\infty} \frac{e^{-2\alpha(\mathbf{r}\pm\mathbf{R})^2}}{|\mathbf{r}\pm\mathbf{R}|}\,\text{d}\mathbf{r} = 
\int^{\infty}_{-\infty} \frac{e^{-2\alpha r^2}}{r}\,\text{d}\mathbf{r} =
4\pi\int^{\infty}_{0} r e^{-2\alpha r^2} \, \text{d}r=\frac{\pi}{\alpha},
\end{equation}
and thus,
\begin{equation}
\boxed{V_{\mu\mu}(A) = V_{\nu\nu}(B) =-2\sqrt{\frac{2\alpha}{\pi}}.}
\end{equation}

\noindent\textbf{Coulomb Integral.} Next is the case where the Gaussians are centered on the same atom, while the operator is centered on the other one,
\begin{equation}
V_{\mu\mu}(B) = V_{\nu\nu}(A) = -\left(\frac{2\alpha}{\pi}\right)^{\frac{3}{2}}
\int^{\infty}_{-\infty} \frac{e^{-2\alpha(\mathbf{r}\pm\mathbf{R})^2}}{|\mathbf{r}\mp\mathbf{R}|}\,\text{d}\mathbf{r}.
\end{equation}
Here, changing the coordinates will not remove the coordinate shift $\pm\mathbf{R}$. One way to make the Coulomb operator more manageable is to use Eq.~\eqref{Cou_to_Gau},
\begin{align}
\int^{\infty}_{-\infty} \frac{e^{-2\alpha(\mathbf{r}\pm\mathbf{R})^2}}{|\mathbf{r}\mp\mathbf{R}|}\,\text{d}\mathbf{r} &= 
\frac{2}{\sqrt{\pi}}\int^{\infty}_{0} \int^{\infty}_{-\infty} e^{-2\alpha(\mathbf{r}\pm\mathbf{R})^2} e^{-t^2(\mathbf{r}\mp\mathbf{R})^2}\,\text{d}\mathbf{r}\,\text{d}t  \nonumber \\
&=\frac{2}{\sqrt{\pi}}\int^{\infty}_{0} e^{-\frac{8\alpha R^2 t^2}{2\alpha+t^2}} \int^{\infty}_{-\infty} e^{-(2\alpha+t^2)(\mathbf{r}-\mathbf{P})^2}\,\text{d}\mathbf{r}\,\text{d}t,
\end{align}
where $\mathbf{P}$ is a point on the $z$-axis (see Eq.~\eqref{GPT}) which can be removed by substitution,
\begin{equation}
\frac{2}{\sqrt{\pi}}\int^{\infty}_{0} e^{-\frac{8\alpha R^2 t^2}{2\alpha+t^2}} \int^{\infty}_{-\infty} e^{-(2\alpha+t^2)r^2}\,\text{d}\mathbf{r}\,\text{d}t=2\pi
\int^{\infty}_{0}\frac{1}{(2\alpha+t^2)^{\frac{3}{2}}} e^{-\frac{8\alpha R^2 t^2}{2\alpha+t^2}}\,\text{d}t.
\label{Vinter}
\end{equation}
We will now apply the substitution in Eq.~\eqref{ISubs} and use Eq.~\eqref{IErf},
\begin{equation}
2\pi
\int^{\infty}_{0}\frac{1}{(2\alpha+t^2)^{\frac{3}{2}}} e^{-\frac{8\alpha R^2 t^2}{2\alpha+t^2}}\,\text{d}t =
\frac{\pi}{\alpha}\int^{1}_{0} e^{-8\alpha R^2 u^2}\, \text{d}u=\left(\frac{\pi}{2\alpha}\right)^{\frac{3}{2}}\frac{\text{erf}(2\sqrt{2\alpha}R)}{2R}.
\end{equation}
Thus, eventually,
\begin{equation}
\boxed{V_{\mu\mu}(B) = V_{\nu\nu}(A) = -\frac{\text{erf}(2\sqrt{2\alpha}R)}{2R}.}
\end{equation}

\noindent\textbf{Resonance Integral.} Finally, the case where the two Gaussians are centered on different atoms,
\begin{equation}
V_{\mu\nu}(A) = V_{\mu\nu}(B) = -\left(\frac{2\alpha}{\pi}\right)^{\frac{3}{2}} e^{-2\alpha R^2}
\int^{\infty}_{-\infty} \frac{e^{-2\alpha \mathbf{r}^2}}{|\mathbf{r}\pm\mathbf{R}|}\,\text{d}\mathbf{r}.
\end{equation}
The evaluation follows closely the previous case except that here we have a factor of $-2\alpha R^2$ in the exponent instead of the factor $-8\alpha R^2$ in Eq.~\eqref{Vinter},
\begin{equation}
\frac{2}{\sqrt{\pi}}\int^{\infty}_{0} e^{-\frac{2\alpha R^2 t^2}{2\alpha+t^2}} \int^{\infty}_{-\infty} e^{-(2\alpha+t^2)(\mathbf{r}-\mathbf{P})^2}\,\text{d}\mathbf{r}\,\text{d}t
=\frac{\pi}{\alpha}
\int^{1}_{0} e^{-2\alpha R^2 u^2}\, \text{d}u=\left(\frac{\pi}{2\alpha}\right)^{\frac{3}{2}}\frac{\text{erf}(\sqrt{2\alpha}R)}{R},
\end{equation}
which finally leads to
\begin{equation}
\boxed{V_{\mu\nu}(A) = V_{\mu\nu}(B) =-\frac{\text{erf}(\sqrt{2\alpha}R)}{R}e^{-2\alpha R^2}.}
\end{equation}

\subsection{Electron-Electron Repulsion Integrals}

\noindent\textbf{Atomic Integral.} The two-body terms can be dealt with similarly.  Since there are only two basis functions, there are only four unique integrals. In the first of these, all basis functions are on the same center,
\begin{equation}
(\mu\mu|\mu\mu) = (\nu\nu|\nu\nu) = \left(\frac{2\alpha}{\pi}\right)^3 \int^{\infty}_{-\infty}\int^{\infty}_{-\infty} \frac{e^{-2\alpha(\mathbf{r}_1\pm\mathbf{R})^2}e^{-2\alpha(\mathbf{r}_2\pm\mathbf{R})^2}}{|\mathbf{r}_1 - \mathbf{r}_2|} \,\text{d}\mathbf{r}_1\text{d}\mathbf{r}_2,
\end{equation}
and because $|\mathbf{r}_1 - \mathbf{r}_2| = |(\mathbf{r}_1\pm\mathbf{R}) - (\mathbf{r}_2\pm\mathbf{R})|$, this integral can be simply reduced to
\begin{align}
&\int^{\infty}_{-\infty}\int^{\infty}_{-\infty} \frac{e^{-2\alpha\mathbf{r}_1^2}e^{-2\alpha\mathbf{r}_2^2}}{|\mathbf{r}_1 - \mathbf{r}_2|} \,\text{d}\mathbf{r}_1\text{d}\mathbf{r}_2 =
\nonumber \\
&\frac{2}{\sqrt{\pi}}\int^{\infty}_{0}\int^{\infty}_{-\infty}\int^{\infty}_{-\infty} e^{-2\alpha\mathbf{r}_1^2} e^{-t^2|\mathbf{r}_1 - \mathbf{r}_2|^2} e^{-2\alpha\mathbf{r}_2^2} \,\text{d}\mathbf{r}_1\text{d}\mathbf{r}_2\text{d}t.
\label{geri1}
\end{align}
By a simple application of Eq.~\eqref{GPT}, the integral over $\mathbf{r}_1$ becomes
\begin{align}
\int^{\infty}_{-\infty} e^{-2\alpha\mathbf{r}_1^2} e^{-t^2|\mathbf{r}_1 - \mathbf{r}_2|^2} \,\text{d}\mathbf{r}_1 &=
e^{-\frac{2\alpha t^2 \mathbf{r}_2^2}{2\alpha+t^2}} \int^{\infty}_{-\infty} e^{-(2\alpha+t^2)(\mathbf{r}_1 - \mathbf{P})^2} \,\text{d}\mathbf{r}_1 \nonumber \\
&=\left(\frac{\pi}{2\alpha + t^2}\right)^{\frac{3}{2}} e^{-\frac{2\alpha t^2 \mathbf{r}_2^2}{2\alpha+t^2}},
\label{ExpCou1}
\end{align}
where $\mathbf{P}$ now does not necessarily lie on the $z$-axis and it depends on $x_2$, $y_2$ and $z_2$, but it is still just a shift as far as the integration with respect to the first set of coordinates is concerned.  Substituting this back into Eq.~\eqref{geri1},
\begin{align}
&2\pi\int^{\infty}_{0}\frac{1}{(2\alpha+t^2)^{\frac{3}{2}}}\int^{\infty}_{-\infty} e^{-\frac{2\alpha t^2 \mathbf{r}_2^2}{2\alpha+t^2}} e^{-2\alpha\mathbf{r}_2^2} \,\text{d}\mathbf{r}_2\text{d}t = 
\nonumber \\
&2\pi\int^{\infty}_{0} \frac{1}{(2\alpha+t^2)^{\frac{3}{2}}} \int^{\infty}_{-\infty} e^{-\frac{4\alpha (\alpha+t^2) \mathbf{r}_2^2}{2\alpha+t^2}} \,\text{d}\mathbf{r}_2\text{d}t.
\label{geri1a}
\end{align}
Integrating with respect to the second set of coordinates leads to the following simple integral over $t$,
\begin{equation}
\frac{\pi}{4}\left(\frac{\pi}{\alpha}\right)^{\frac{3}{2}}\int^{\infty}_{0} \frac{1}{(\alpha+t^2)^{\frac{3}{2}}} \,\text{d}t = \frac{1}{4}\left(\frac{\pi}{\alpha}\right)^{\frac{5}{2}},
\end{equation}
remembering that the integrand is essentially the derivative of $t/\sqrt{\alpha+t^2}$. Multiplication with the norm factor gives
\begin{equation}
\boxed{(\mu\mu|\mu\mu) = (\nu\nu|\nu\nu) = 2\sqrt{\frac{\alpha}{\pi}}.}
\end{equation}

\noindent\textbf{Hybrid Integral.} The evaluation of the next integral,
\begin{equation}
(\mu\mu|\mu\nu) = (\nu\nu|\mu\nu) = \left(\frac{2\alpha}{\pi}\right)^3 e^{-2\alpha R^2} \int^{\infty}_{-\infty}\int^{\infty}_{-\infty} \frac{e^{-2\alpha\mathbf{r}_1^2}e^{-2\alpha(\mathbf{r}_2\pm\mathbf{R})^2}}{|\mathbf{r}_1 - \mathbf{r}_2|} \,\text{d}\mathbf{r}_1\text{d}\mathbf{r}_2,
\end{equation}
is very similar up to the integration over $\mathbf{r}_2$ in Eq.~\eqref{geri1a}, where a second application of Eq.~\eqref{GPT} is called for
\begin{align}
&2\pi\int^{\infty}_{0}\frac{1}{(2\alpha+t^2)^{\frac{3}{2}}}\int^{\infty}_{-\infty} e^{-\frac{2\alpha t^2 \mathbf{r}_2^2}{2\alpha+t^2}} e^{-2\alpha(\mathbf{r}_2\pm\mathbf{R})^2} \,\text{d}\mathbf{r}_2\text{d}t = 
\nonumber \\
& 2\pi\int^{\infty}_{0} \frac{e^{-\frac{\alpha t^2 R^2}{\alpha + t^2}}}{(2\alpha+t^2)^{\frac{3}{2}}} \int^{\infty}_{-\infty} e^{-\frac{4\alpha (\alpha+t^2) (\mathbf{r}_2-\mathbf{Q})^2}{2\alpha+t^2}}\,\text{d}\mathbf{r}_2\text{d}t.
\end{align}
This yields a new point $\mathbf{Q}$ as the center of the resulting product. The exact position need not concern us, however, since it can be removed via a change of variables. Thus,  the integral becomes
\begin{equation}
\frac{\pi}{4}\left(\frac{\pi}{\alpha}\right)^{\frac{3}{2}}\int^{\infty}_{0} \frac{e^{-\frac{\alpha t^2 R^2}{\alpha + t^2}}}{(\alpha+t^2)^{\frac{3}{2}}} \,\text{d}t =
 \frac{1}{4}\left(\frac{\pi}{\alpha}\right)^{\frac{5}{2}}\int^1_0 e^{-\alpha R^2 u^2}\,\text{d}u,
\label{sub_erf_1}
\end{equation}
after the substitution $u^2=t^2/(\alpha+t^2)$ and $\text{d}u=\alpha\text{d}t/(\alpha+t^2)^{3/2}$. Via Eq.~\eqref{IErf}, this yields
\begin{equation}
\boxed{(\mu\mu|\mu\nu) = (\nu\nu|\mu\nu) = \frac{\text{erf}(\sqrt{\alpha}R)}{R}e^{-2\alpha R^2}.}
\end{equation}

\noindent\textbf{Coulomb Integral.} The penultimate integral has the form
\begin{equation}
(\mu\mu|\nu\nu) = (\nu\nu|\mu\mu) = \left(\frac{2\alpha}{\pi}\right)^3 \int^{\infty}_{-\infty}\int^{\infty}_{-\infty} \frac{e^{-2\alpha(\mathbf{r}_1\pm\mathbf{R})^2}e^{-2\alpha(\mathbf{r}_2\mp\mathbf{R})^2}}{|\mathbf{r}_1 - \mathbf{r}_2|} \,\text{d}\mathbf{r}_1\text{d}\mathbf{r}_2,
\end{equation}
The procedure is similar as before in Eq.~\eqref{geri1}, except that when the Coulomb potential is converted into a Gaussian form via Eq.~\eqref{Cou_to_Gau} and multiplied with the Gaussian function depending on the first set of coordinates, this yields via Eq.~\eqref{GPT} a Gaussian depending on $\mathbf{r}_2\pm\mathbf{R}$,
\begin{align}
&\frac{2}{\sqrt{\pi}}\int^{\infty}_{0}\int^{\infty}_{-\infty} e^{-\frac{2\alpha t^2 (\mathbf{r}_2\pm\mathbf{R})^2}{2\alpha+t^2}} \int e^{-(2\alpha+t^2)(\mathbf{r}_1 - \mathbf{P})^2} \,\text{d}\mathbf{r}_1 e^{-2\alpha(\mathbf{r}_2\mp\mathbf{R})^2} \,\text{d}\mathbf{r}_2\text{d}t = \nonumber \\
&2\pi\int^{\infty}_{0}
\frac{e^{-\frac{4\alpha t^2 R^2}{\alpha + t^2}}}{(2\alpha+t^2)^{\frac{3}{2}}}
\int^{\infty}_{-\infty} 
e^{-\frac{4\alpha (\alpha + t^2)}{2\alpha + t^2}(\mathbf{r}_2-\mathbf{Q})^2} \,\text{d}\mathbf{r}_2\text{d}t =
\frac{\pi}{4}\left(\frac{\pi}{\alpha}\right)^{\frac{3}{2}}\int^{\infty}_{0} \frac{e^{-\frac{4\alpha t^2 R^2}{\alpha + t^2}}}{(\alpha+t^2)^{\frac{3}{2}}} \,\text{d}t.
\end{align}
The vector $\mathbf{Q}$ again comes from Eq.~\eqref{GPT}, and is irrelevant to the integration over $\mathbf{r}_2$. The last integral is almost the same as the one in Eq.~\eqref{sub_erf_1}, and evaluates to
\begin{equation}
\boxed{(\mu\mu|\nu\nu) = (\nu\nu|\mu\mu) = \frac{\text{erf}(2\sqrt{\alpha}R)}{2R}.}
\end{equation}

\noindent\textbf{Exchange Integral.} The last integral
\begin{equation}
(\mu\nu|\mu\nu) = (\nu\mu|\nu\mu) = \left(\frac{2\alpha}{\pi}\right)^3 e^{-4\alpha R^2} \int^{\infty}_{-\infty}\int^{\infty}_{-\infty} \frac{e^{-2\alpha \mathbf{r}_1^2}e^{-2\alpha\mathbf{r}_2^2}}{|\mathbf{r}_1 - \mathbf{r}_2|} \,\text{d}\mathbf{r}_1\text{d}\mathbf{r}_2,
\end{equation}
 is simple since it is essentially the same as in Eq.~\eqref{geri1},
\begin{equation}
\boxed{(\mu\nu|\mu\nu) = (\nu\mu|\nu\mu) = 2\sqrt{\frac{\alpha}{\pi}}e^{-4\alpha R^2}.}
\end{equation}

\section{Slater integrals}

\subsection{Overlap Integrals}

\noindent\textbf{Atomic Integral (Norm).} The factor $N_S$ can be found in a similar way as in the case of Gaussians
\begin{equation}
N_S^2\int^{\infty}_{-\infty} e^{-2\alpha|\mathbf{r}\pm\mathbf{R}|}\,\text{d}\mathbf{r} =
N_S^2\int^{\infty}_{-\infty} e^{-2\alpha|\mathbf{r}|}\,\text{d}\mathbf{r} =
4\pi N_S^2\int^{\infty}_{0} r^2 e^{-2\alpha r}\,\text{d}r = N_S^2 \frac{\pi}{\alpha^3}.
\end{equation}
Setting this to one, we find that the normalized Slater orbitals have the form
\begin{equation}
\chi^S_{\mu} = \sqrt{\frac{\alpha^3}{\pi}}e^{-\alpha |\mathbf{r}+\mathbf{R}|},
\quad
\chi^S_{\nu} =  \sqrt{\frac{\alpha^3}{\pi}}e^{-\alpha |\mathbf{r}-\mathbf{R}|}.
\end{equation}
By definition, the atomic overlap integral
\begin{equation}
S_{\mu\mu} = S_{\nu\nu} = \frac{\alpha^3}{\pi}
\int^{\infty}_{-\infty} e^{-2\alpha |\mathbf{r}\pm\mathbf{R}|}\,\text{d}\mathbf{r}
\end{equation}
is then simply
\begin{equation}
\boxed{S_{\mu\mu} = S_{\nu\nu} = 1.}
\label{OvlS}
\end{equation}

\noindent\textbf{Resonance Integral.} For two different functions, we may use prolate ellipsoidal coordinates in the form given in Eq.~\eqref{PECnoAngle},
\begin{equation}
S_{\mu\nu} = S_{\nu\mu} = \frac{\alpha^3}{\pi}
\int^{\infty}_{-\infty} e^{-\alpha|\mathbf{r}\pm\mathbf{R}|}e^{-\alpha|\mathbf{r}\mp\mathbf{R}|}\,\text{d}\mathbf{r} =
2\alpha^3 R^3\int^{\infty}_{1}\int^{1}_{-1} e^{-2\alpha R\sigma}(\sigma^2-\tau^2)\,\text{d}\tau\text{d}\sigma,
\end{equation}
and, after the integration with respect to $\tau$,
\begin{equation}
4\alpha^3 R^3\int^{\infty}_{1} e^{-2\alpha R\sigma}\left(\sigma^2-\frac{1}{3}\right)\,\text{d}\sigma.
\end{equation}
From Eq.~\eqref{Spoly}, this is then
\begin{equation}
\boxed{S_{\mu\nu} = S_{\nu\mu} = e^{-2\alpha R}\left(\frac{4\alpha^2 R^2}{3}+2\alpha R+1\right).}
\label{ssmn}
\end{equation}

\subsection{Kinetic Integrals}

\noindent\textbf{Atomic Integral.} The first kinetic integral takes the form
\begin{equation}
T_{\mu\mu}=T_{\nu\nu} = -\frac{1}{2}\frac{\alpha^3}{\pi} \int^{\infty}_{-\infty} e^{-\alpha|\mathbf{r}\pm\mathbf{R}|}\nabla^2 e^{-\alpha|\mathbf{r}\pm\mathbf{R}|}\,\text{d}\mathbf{r}.
\end{equation}
The gradient in this case is $\nabla\chi (\mathbf{r}_{\pm}) = -\alpha\mathbf{r}_{\pm}\chi (\mathbf{r}_{\pm})/r_{\pm}$, and the integral becomes
\begin{equation}
\int^{\infty}_{-\infty} \nabla e^{-\alpha|\mathbf{r}\pm\mathbf{R}|}\cdot \nabla e^{-\alpha|\mathbf{r}\pm\mathbf{R}|}\,\text{d}\mathbf{r}
=\alpha^2 \int^{\infty}_{-\infty} e^{-2\alpha r_{\pm}} \,\text{d}\mathbf{r},
\end{equation}
which is basically the overlap integral in Eq.~\eqref{OvlS}. Thus,
\begin{equation}
\boxed{T_{\mu\mu}=T_{\nu\nu} = \frac{\alpha^2}{2}.}
\end{equation}

\noindent\textbf{Resonance Integral.} The other kinetic integral is
\begin{equation}
T_{\mu\nu}=T_{\nu\mu} = -\frac{1}{2}\frac{\alpha^3}{\pi} \int^{\infty}_{-\infty} e^{-\alpha|\mathbf{r}\pm\mathbf{R}|}\nabla^2 e^{-\alpha|\mathbf{r}\mp\mathbf{R}|}\,\text{d}\mathbf{r}.
\end{equation}
Continuing as before, via Eq.~\eqref{IParts},
\begin{equation}
\int^{\infty}_{-\infty} \nabla e^{-\alpha|\mathbf{r}\pm\mathbf{R}|}\cdot \nabla e^{-\alpha|\mathbf{r}\mp\mathbf{R}|}\,\text{d}\mathbf{r} =
\alpha^2\int^{\infty}_{-\infty} \frac{e^{-\alpha r_\pm} e^{-\alpha r_\mp}}{r_\pm r_\mp}(r^2-R^2)\,\text{d}\mathbf{r},
\end{equation}
and switching to prolate spheroidal coordinates, via Eq.~\eqref{PECnoAngle}, we get
\begin{equation}
2\pi\alpha^2 R^3 \int^{\infty}_{1} e^{-2\alpha R\sigma} \int^{1}_{-1}  (\sigma^2+\tau^2-2)\,\text{d}\tau \text{d}\sigma =
4\pi\alpha^2 R^3 \int^{\infty}_{1} e^{-2\alpha R\sigma} (\sigma^2-\frac{5}{3})\, \text{d}\sigma.
\end{equation}
The final result is thus
\begin{equation}
\boxed{T_{\mu\nu} = T_{\nu\mu} = -\frac{1}{2}\alpha^2 e^{-2\alpha R}\left(\frac{4\alpha^2 R^2}{3}-2\alpha R-1\right).}
\end{equation}

\subsection{Nuclear-Electron Attraction Integrals}

\noindent\textbf{Atomic Integral.} The first integral is
\begin{equation}
V_{\mu\mu}(A) = V_{\nu\nu}(B) = -\frac{\alpha^3}{\pi}
\int^{\infty}_{-\infty} \frac{e^{-2\alpha |\mathbf{r}\pm\mathbf{R}|}}{|\mathbf{r}\pm\mathbf{R}|}\,\text{d}\mathbf{r},
\end{equation}
which is most easily evaluated in spherical coordinates, as in Eq.~\eqref{SPCnoAngle},
\begin{equation}
\int^{\infty}_{-\infty} \frac{e^{-2\alpha r}}{r}\,\text{d}\mathbf{r} =
4\pi\int^{\infty}_{0} r e^{-2\alpha r} \, \text{d}r=\frac{\pi}{\alpha^2},
\end{equation}
where the last step follows from Eq.~\eqref{Spoly}. Thus,
\begin{equation}
\boxed{V_{\mu\mu}(A) = V_{\nu\nu}(B) = -\alpha.}
\end{equation}

\noindent\textbf{Coulomb Integral.} Next, we have
\begin{equation}
V_{\mu\mu}(B) = V_{\nu\nu}(A) = -\frac{\alpha^3}{\pi}
\int^{\infty}_{-\infty} \frac{e^{-2\alpha |\mathbf{r}\pm\mathbf{R}|}}{|\mathbf{r}\mp\mathbf{R}|}\,\text{d}\mathbf{r},
\label{svmm}
\end{equation}
which takes the form, by Eq.~\eqref{PECnoAngle}, in prolate spheroidal coordinates
\begin{equation}
\int^{\infty}_{-\infty} \frac{e^{-2\alpha r_{\pm}}}{r_{\mp}}\,\text{d}\mathbf{r} =
2\pi R^2\int^{\infty}_{1}\int^{1}_{-1} e^{-2\alpha R(\sigma\pm\tau)}(\sigma\pm\tau) \, \text{d}\tau\, \text{d}\sigma.
\end{equation}
This may be factorized as follows
\begin{equation}
2\pi R^2
\int^{\infty}_{1} e^{-2\alpha R\sigma}\sigma \, \text{d}\sigma
\int^{1}_{-1}  e^{\mp 2\alpha R\tau} \, \text{d}\tau \pm
2\pi R^2
\int^{\infty}_{1} e^{-2\alpha R\sigma} \, \text{d}\sigma
\int^{1}_{-1} e^{\mp 2\alpha R\tau}\tau \, \text{d}\tau
\end{equation}
These integrals can be easily evaluated using Eq.~\eqref{Spoly},
\begin{equation}
\int^{\infty}_{-\infty} \frac{e^{-2\alpha r_{\pm}}}{r_{\mp}}\,\text{d}\mathbf{r} =\frac{\pi}{2\alpha^3 R}\left(1-(1+2\alpha R)e^{-4\alpha R}\right),
\end{equation}
and thus
\begin{equation}
\boxed{V_{\mu\mu}(B) = V_{\nu\nu}(A) = 
-\frac{1}{2R}\left(1-(1+2\alpha R)e^{-4\alpha R}\right).}
\end{equation}

\noindent\textbf{Resonance Integral.} The last integral of this type is
\begin{equation}
V_{\mu\nu}(A) = V_{\mu\nu}(B) = -\frac{\alpha^3}{\pi}
\int^{\infty}_{-\infty} \frac{e^{-\alpha |\mathbf{r}\pm\mathbf{R}|}e^{-\alpha |\mathbf{r}\mp\mathbf{R}|}}{|\mathbf{r}\pm\mathbf{R}|}\,\text{d}\mathbf{r},
\label{svmn}
\end{equation}
which in prolate spheroidal coordinates reads
\begin{equation}
\int^{\infty}_{-\infty} \frac{e^{-\alpha r_\pm}e^{-\alpha r_\mp}}{r_\pm}\,\text{d}\mathbf{r} =
2\pi R^2\int^{\infty}_{1}\int^{1}_{-1} e^{-2\alpha R\sigma}(\sigma\pm\tau) \, \text{d}\tau\, \text{d}\sigma =
4\pi R^2\int^{\infty}_{1} e^{-2\alpha R\sigma}\sigma \, \text{d}\sigma.
\end{equation}
The last integral is simply evaluated using Eq.~\eqref{Spoly}, and thus the final result is
\begin{equation}
\boxed{V_{\mu\nu}(A) = V_{\mu\nu}(B) = -\alpha(1+2\alpha R)e^{-2\alpha R}.}
\end{equation}

\subsection{Electron-Electron Repulsion Integrals}

\noindent\textbf{Atomic Integral.} The simplest of the repulsion integrals,
\begin{equation}
(\mu\mu|\mu\mu) = (\nu\nu|\nu\nu) = \frac{\alpha^6}{\pi^2} \int^{\infty}_{-\infty}\int^{\infty}_{-\infty} \frac{e^{-2\alpha |\mathbf{r}_1\pm\mathbf{R}|}e^{-2\alpha |\mathbf{r}_2\pm\mathbf{R}|}}{|\mathbf{r}_1 - \mathbf{r}_2|} \,\text{d}\mathbf{r}_1\text{d}\mathbf{r}_2,
\end{equation}
can be dealt with by a simple application of Eq.~\eqref{SExp2} with the choice $\tilde{f}(r_1,r_2)=\alpha^6 e^{-2\alpha (r_1+r_2)}/\pi^2$, and noting that $\tilde{f}(r_1,r_2)=\tilde{f}(r_2,r_1)$,
\begin{equation}
(\mu\mu|\mu\mu) =
32\alpha^6\int^\infty_0 r_1 e^{-2\alpha r_1}\int^{r_1}_0 r_2^2 e^{-2\alpha r_2}\,\text{d}r_2\,\text{d}r_1,
\end{equation}
which gives
\begin{equation}
\boxed{(\mu\mu|\mu\mu) = (\nu\nu|\nu\nu) = \frac{5}{8}\alpha.}
\end{equation}

\noindent\textbf{Hybrid Integral.} In this two-center integral, only one of the four possible orbitals are located on a different center than the others,
\begin{equation}
(\mu\mu|\mu\nu) = (\nu\nu|\mu\nu) =  \frac{\alpha^6}{\pi^2} \int^{\infty}_{-\infty}\int^{\infty}_{-\infty} \frac{e^{-2\alpha|\mathbf{r}_1\pm \mathbf{R}|}e^{-\alpha|\mathbf{r}_2\pm\mathbf{R}|}e^{-\alpha|\mathbf{r}_2\mp\mathbf{R}|}}{|\mathbf{r}_1 - \mathbf{r}_2|} \,\text{d}\mathbf{r}_1\text{d}\mathbf{r}_2.
\end{equation}
However, this integral still has two identical orbitals describing the same electron. One can deal with such integrals by reducing them to one electron integrals by integrating over the first set of coordinates. Thus, let us convert all quantities into the Gaussian basis by Eq.~\eqref{STO_to_GTO} and Eq.~\eqref{Cou_to_Gau},
\begin{equation}
\int^{\infty}_{-\infty} \frac{e^{-2\alpha|\mathbf{r}_1\pm \mathbf{R}|}}{|\mathbf{r}_1 - \mathbf{r}_2|} \,\text{d}\mathbf{r}_1 =\frac{2\alpha}{\pi}
\int^{\infty}_{0} \frac{1}{\sqrt{\xi^3}} e^{-\frac{\alpha^2}{\xi}} 
\int^{\infty}_{0} \int^{\infty}_{-\infty} e^{-\xi r_{1_\pm}^2} e^{-r_{12}^2 t^2}\,\text{d}\mathbf{r}_1\,\text{d}t\,\text{d}\xi.
\end{equation}
We recognize the integral over $\mathbf{r}_1$ from Eq.~\eqref{ExpCou1},
\begin{equation}
2\alpha\sqrt{\pi}
\int^{\infty}_{0} \frac{1}{\sqrt{\xi^3}} e^{-\frac{\alpha^2}{\xi}} 
\int^{\infty}_{0}\frac{1}{(2\alpha + t^2)^{\frac{3}{2}}} e^{-\frac{2\alpha t^2 \mathbf{r}_2^2}{2\alpha+t^2}}\,\text{d}t\,\text{d}\xi =
\frac{\alpha\pi}{r_{2_\pm}}
\int^{\infty}_{0} \frac{1}{\xi^3} e^{-\frac{\alpha^2}{\xi}}
\text{erf}(\sqrt{\xi}r_{2_\pm})
\,\text{d}\xi.
\end{equation}
The integration over $t$ proceeds by substitution as in Eq.~\eqref{ISubs} and gives and error function via Eq.~\eqref{IErf}. The last integral can be dealt with via a change of variables $u=1/\sqrt{\xi}$, $\text{d}u=-\text{d}t/(2\sqrt{\xi^3})$ and the appropriate transformation of the integration limits,
\begin{align}
\int^{\infty}_{0} \frac{1}{\xi^3} e^{-\frac{\alpha^2}{\xi}}
\text{erf}(\sqrt{\xi}r_{2_\pm})
\,\text{d}\xi &= 
2\int^{\infty}_{0} u^3 e^{-\alpha^2 u^2}
\text{erf}(r_{2_\pm}/u)
\,\text{d}u.
\end{align}
This integral can be easily evaluated using integration by parts from Eq.~\eqref{GauX1X} and Eq.~\eqref{xErfGau}. Thus, eventually,
\begin{equation}
\int^{\infty}_{-\infty} \frac{e^{-2\alpha|\mathbf{r}_1\pm \mathbf{R}|}}{|\mathbf{r}_1 - \mathbf{r}_2|} \,\text{d}\mathbf{r}_1 =\frac{\pi}{\alpha^3}\frac{1-e^{-2\alpha r_{2_\pm}}(1+\alpha r_{2_\pm})}{r_{2_\pm}}.
\label{Rscreened}
\end{equation}
On substitution, the original integral thus becomes
\begin{align}
(\mu\mu|\mu\nu) 
&= \frac{\alpha^3}{\pi} \int^{\infty}_{-\infty}\frac{e^{-\alpha (r_{2_\pm}+r_{2_\mp})}}{r_{2_\pm}}\,\text{d}\mathbf{r}_2
\nonumber \\
&- \frac{\alpha^3}{\pi} \int^{\infty}_{-\infty}\frac{e^{-\alpha (3r_{2_\pm}+r_{2_\mp})}}{r_{2_\pm}}\,\text{d}\mathbf{r}_2
\nonumber \\
&- \frac{\alpha^4}{\pi} \int^{\infty}_{-\infty}e^{-\alpha (3r_{2_\pm}+r_{2_\mp})}\,\text{d}\mathbf{r}_2.
\end{align}
The first term is the same as Eq.~\eqref{svmn} apart from a sign. The second term is a slightly different integral that differs only in the exponent. In prolate ellipsoidal coordinates, as in Eq.~\eqref{PECnoAngle},
\begin{align}
\int^{\infty}_{-\infty}\frac{e^{-\alpha (3r_{2_\pm}+r_{2_\mp})}}{r_{2_\pm}}\,\text{d}\mathbf{r}_2 &= 2\pi R^2 \int^{\infty}_{1}\int^{1}_{-1} (\sigma\mp\tau) e^{-2\alpha R (2\sigma\pm\tau)}\,\text{d}\tau\,\text{d}\sigma
\nonumber \\
&=\frac{\pi}{16\alpha^3 R} e^{-6\alpha R}(1+e^{4\alpha R}(8\alpha R - 1)).
\end{align}
Similarly, the last term is closely related to Eq.~\eqref{ssmn},
\begin{align}
\int^{\infty}_{-\infty} e^{-\alpha (3r_{2_\pm}+r_{2_\mp})}\,\text{d}\mathbf{r}_2 &= 2\pi R^3 \int^{\infty}_{1}\int^{1}_{-1} (\sigma^2-\tau^2) e^{-2\alpha R (2\sigma\pm\tau)}\,\text{d}\tau\,\text{d}\sigma
\nonumber \\
&=\frac{\pi}{32\alpha^4 R} e^{-6\alpha R}(3+4\alpha R+3e^{4\alpha R}(4\alpha R - 1)).
\end{align}
Thus, we have eventually,
\begin{equation}
\boxed{
(\mu\mu|\mu\nu) = (\nu\nu|\mu\nu) = 
-\frac{1}{32 R} \left(5+4\alpha R-(5+4\alpha R (1+16\alpha R))e^{4\alpha R}\right)e^{-6\alpha R}.}
\end{equation}

\noindent\textbf{Coulomb Integral.} The evaluation of the Coulomb integral
\begin{equation}
(\mu\mu|\nu\nu) = (\nu\nu|\mu\mu) =  \frac{\alpha^6}{\pi^2} \int^{\infty}_{-\infty}\int^{\infty}_{-\infty} \frac{e^{-2\alpha|\mathbf{r}_1\pm \mathbf{R}|}e^{-2\alpha|\mathbf{r}_2\mp\mathbf{R}|}}{|\mathbf{r}_1 - \mathbf{r}_2|} \,\text{d}\mathbf{r}_1\text{d}\mathbf{r}_2,
\end{equation}
also proceeds via Eq.~\eqref{Rscreened}. Upon substitution,
\begin{align}
(\mu\mu|\nu\nu) 
&= \frac{\alpha^3}{\pi} \int^{\infty}_{-\infty}\frac{e^{-2\alpha r_{2_\mp}}}{r_{2_\pm}}\,\text{d}\mathbf{r}_2
\nonumber \\
&- \frac{\alpha^3}{\pi} \int^{\infty}_{-\infty}\frac{e^{-2\alpha (r_{2_\pm}+r_{2_\mp})}}{r_{2_\pm}}\,\text{d}\mathbf{r}_2
\nonumber \\
&- \frac{\alpha^4}{\pi} \int^{\infty}_{-\infty}e^{-2\alpha (r_{2_\pm}+r_{2_\mp})}\,\text{d}\mathbf{r}_2.
\end{align}
These integrals can be related to those defined in Eq.~\eqref{svmm}, Eq.~\eqref{svmn}, and Eq.~\eqref{ssmn}, respectively. Thus, the final result is
\begin{equation}
\boxed{
(\mu\mu|\nu\nu) = (\nu\nu|\mu\mu) = 
\frac{1}{24 R} \left(12-(12+\alpha R (33+4\alpha R (9+4\alpha R)))e^{-4\alpha R}\right).}
\end{equation}

\noindent\textbf{Exchange Integral.} The exchange integral was historically the last to be evaluated for the hydrogen molecule,
\begin{equation}
(\mu\nu|\mu\nu) = (\nu\mu|\nu\mu) = \frac{\alpha^3}{\pi} \int^{\infty}_{-\infty}\int^{\infty}_{-\infty} \frac{e^{-\alpha (r_{1_\pm}+r_{1_\mp})}e^{-\alpha (r_{2_\pm}+r_{2_\mp})}}{|\mathbf{r}_1 - \mathbf{r}_2|} \,\text{d}\mathbf{r}_1\text{d}\mathbf{r}_2.
\end{equation}
We will follow the techinque used when it was first calculated via the Neumann expansion. Essentially, this involves evaluating Eq.~\eqref{NeuSigma} for the case $\tilde{f}(\sigma_1,\sigma_2)=e^{2\alpha R(\sigma_1+\sigma_2)}$, 
\begin{align}
(\mu\nu|\mu\nu)
&=\frac{16}{15}\alpha^6 R^5 
\int^\infty_1 \ln\frac{\sigma_1+1}{\sigma_1-1} e^{-2\alpha R\sigma_1} (3\sigma_1^2-1) \int^{\sigma_1}_1 e^{-2\alpha R\sigma_2} (3\sigma_2^2-1) \,\text{d}\sigma_2\text{d}\sigma_1 \nonumber\\
&+\frac{16}{15}\alpha^6 R^5
\int^\infty_1 e^{-2\alpha R\sigma_1} (3\sigma_1^2-1) \int^\infty_{\sigma_1}\ln\frac{\sigma_2+1}{\sigma_2-1} e^{-2\alpha R\sigma_2} (3\sigma_2^2-1) \,\text{d}\sigma_2\text{d}\sigma_1 \nonumber\\
&-\frac{16}{15}\alpha^6 R^5
\int^\infty_1 \sigma_1 e^{-2\alpha R\sigma_1} \int^{\sigma_1}_1 e^{-2\alpha R\sigma_2} (3\sigma_2^2-1) \,\text{d}\sigma_2\text{d}\sigma_1 \nonumber\\
&-\frac{16}{15}\alpha^6 R^5
\int^\infty_1 e^{-2\alpha R\sigma_1} (3\sigma_1^2-1) \int^\infty_{\sigma_1} \sigma_2 e^{-2\alpha R\sigma_2} \,\text{d}\sigma_2\text{d}\sigma_1.
\label{NforEx}
\end{align}
Most of the integrals involved are elementary, if somewhat tedious to evaluate. It will be useful to introduce the following intermediate,
\begin{equation}
S_\pm = (3\pm 2\alpha R(3\pm 2\alpha R))e^{\mp 2\alpha R}.
\end{equation}
Turning to the integrals over $\sigma_2$ first, neither
\begin{equation}
\int^{\infty}_{\sigma_1} \sigma_2 e^{-2\alpha R\sigma_2} \,\text{d}\sigma_2=
\frac{1}{4\alpha^2 R^2}(1+2\alpha R\sigma_1)e^{-2\alpha R\sigma_1},
\end{equation}
nor
\begin{equation}
\int^{\sigma_1}_1 e^{-2\alpha R\sigma_2} (3\sigma_2^2-1) \,\text{d}\sigma_2=
\frac{1}{4\alpha^3 R^3} \left(S_+ - (3 + 2\alpha R(3\sigma_1 +\alpha R(3\sigma_1^2-1)))e^{2\alpha R\sigma_1}\right)
\end{equation}
present much difficulty. The integrals containing logarithms require more attention.  The simplest of these has the form
\begin{equation}
\int^\infty_{\sigma_1} \ln (\sigma_2-1) e^{-2\alpha R\sigma_2}\,\text{d}\sigma_2 =
\frac{1}{2\alpha R}\left(e^{-2\alpha R\sigma_1}\ln (\sigma_1-1)-e^{-2\alpha R}\text{Ei}(-2\alpha R(\sigma_1-1))\right),
\label{simpleI}
\end{equation}
and can be evaluated from Eq.~\eqref{IExpLn}, keeping in mind the definition of the exponential integral given in Eq.~\eqref{ExpI}. The other integrals can be evaluated using integration by parts, and thus the final integral over $\sigma_2$ is
\begin{align}
&\int^\infty_{\sigma_1}\ln\frac{\sigma_2+1}{\sigma_2-1} e^{-2\alpha R\sigma_2} (3\sigma_2^2-1) \,\text{d}\sigma_2 = \nonumber \\
&\frac{1}{4\alpha^3 R^3} \biggl[
S_+ \text{Ei}(-2\alpha R(\sigma_1 - 1)) - S_- \text{Ei}(-2\alpha R(\sigma_1 + 1))
\nonumber \\
&-\left(6\alpha R - (3 +2\alpha R(3\sigma_1+\alpha R(3\sigma_1^2-1)))\ln\frac{\sigma_1+1}{\sigma_1-1}\right) e^{-2\alpha R\sigma_1}\biggr].
\end{align}
The integration over $\sigma_1$ proceeds along similar lines. The only slight novelty is the appearance of the Euler's constant via Eq.~\eqref{EulerGamma}. Taking an example similar to Eq.~\eqref{simpleI}, the constant appears as the lower integration limit approaches $1$,
\begin{equation}
\int^\infty_{1} \ln (\sigma_1-1) e^{-2\alpha R\sigma_1}\,\text{d}\sigma_1 =
-\frac{1}{2\alpha R}e^{-2\alpha R}\left(\gamma + \ln (2\alpha R) \right).
\end{equation}
Again, the slightly more complicated integrals can be obtained by standard techniques. The fact that the first two and the last two terms in Eq.~\eqref{NforEx} yield identical results simplifies our presentation somewhat. Since $\tilde{f}(\sigma_1,\sigma_2)=\tilde{f}(\sigma_2,\sigma_1)$, two of the four terms is enough, but it will be useful to make this explicit below. The first two terms give
\begin{align}
&\int^\infty_1 \ln\frac{\sigma_1+1}{\sigma_1-1} e^{-2\alpha R\sigma_1} (3\sigma_1^2-1) \int^{\sigma_1}_1 e^{-2\alpha R\sigma_2} (3\sigma_2^2-1) \,\text{d}\sigma_2\text{d}\sigma_1 = \nonumber\\
&
\int^\infty_1 e^{-2\alpha R\sigma_1} (3\sigma_1^2-1) \int^\infty_{\sigma_1}\ln\frac{\sigma_2+1}{\sigma_2-1} e^{-2\alpha R\sigma_2} (3\sigma_2^2-1) \,\text{d}\sigma_2\text{d}\sigma_1 =
\nonumber\\
& \frac{1}{128\alpha^6 R^6}\biggl[
4S_+^2(\gamma+\ln(2\alpha R)) +4S_-^2\text{Ei}(-8\alpha R) -8S_+ S_- \text{Ei}(-4\alpha R)
\nonumber \\
&+3\alpha R (15-4\alpha R(9+8\alpha R)) e^{-4\alpha R} 
\biggr],
\end{align}
while the second two terms lead to
\begin{align}
&\int^\infty_1 \sigma_1 e^{-2\alpha R\sigma_1} \int^{\sigma_1}_1 e^{-2\alpha R\sigma_2} (3\sigma_2^2-1) \,\text{d}\sigma_2\text{d}\sigma_1 =\nonumber\\
&\int^\infty_1 e^{-2\alpha R\sigma_1} (3\sigma_1^2-1) \int^\infty_{\sigma_1} \sigma_2 e^{-2\alpha R\sigma_2} \,\text{d}\sigma_2\text{d}\sigma_1 =\nonumber\\
&\frac{1}{256\alpha^5 R^5} \left(
15+4\alpha R (15+8\alpha R(3+2\alpha R))
\right) e^{-4\alpha R}.
\end{align}
Putting all these terms together and multiplying with the appropriate constant finally gives
\begin{empheq}[box=\fbox]{align}
(\mu\nu|\mu\nu) &=(\nu\mu|\nu\mu) = 
\frac{\alpha}{120}(75-276\alpha R - 288\alpha^2 R^2 -64\alpha^3 R^3)e^{-4\alpha R}
\nonumber \\
+&\frac{1}{15R} \left( S_+^2 (\gamma + \ln (2\alpha R)) + S_-^2 \text{Ei}(-8\alpha R) -2S_+S_- \text{Ei}(-4\alpha R)\right).
\end{empheq}

\end{document}